\documentclass{aa}  
\usepackage{graphicx}
\usepackage{txfonts}
\usepackage{subcaption}
\usepackage{orcidlink}
\usepackage{placeins}
\hypersetup{hidelinks}
\usepackage{natbib}                                     
\usepackage{lscape}                                                 
\begin{document}

   \title{A Brush with Turbulence: Probing the ICM Dynamics in the Merging Toothbrush Cluster using XRISM}

   \author{A. Madangarli\inst{1}\orcidlink{0009-0001-5254-8014}
    \and A. Simionescu\inst{2,3,4}\orcidlink{0000-0002-9714-3862}
    \and H.\,J.\,A. Röttgering\inst{3}\orcidlink{0000-0001-8887-2257}
    \and A. Majumder\inst{5,2}\orcidlink{0000-0002-3525-7186}
    \and S. Haruse \inst{6}
    \and K. Nakazawa \inst{7,6}\orcidlink{0000-0003-2930-350X} 
    \and Y. Uchida \inst{8}\orcidlink{0000-0002-7962-4136}
    \and D. Ito \inst{6}\orcidlink{0009-0000-4742-5098}
    \and G. Brunetti\inst{9}\orcidlink{0000-0003-4195-8613}
    \and R.\,J. van Weeren\inst{3}\orcidlink{0000-0002-0587-1660}
    \and C. Zhang\inst{10,11}\orcidlink{0000-0001-5888-7052}
    \and K. Rajpurohit\inst{12}\orcidlink{0000-0001-7509-2972}
    \and P. Chakraborty\inst{12,13}\orcidlink{0000-0002-4469-2518}
    \and H. Akamatsu\inst{14,2}\orcidlink{0000-0003-1949-7005}}

   \institute{Lorentz Institute for Theoretical Physics, Leiden University, PO Box 9506, NL-2300 RA Leiden, The Netherlands
   \and SRON, Space Research Organisation Netherlands, Niels Bohrweg 4, 2333 CA Leiden, The Netherlands
   \and Leiden Observatory, Leiden University, PO Box 9513, 2300 RA Leiden, The Netherlands
   \and Kavli Institute for the Physics and Mathematics of the Universe (WPI), The University of Tokyo, Kashiwa, Chiba 277-8583, Japan
   \and  Waterloo Centre for Astrophysics, Department of Physics and Astronomy, University of Waterloo, 200 University Avenue West, Waterloo, Ontario N2L 3G1, Canada
   \and Department of Physics, Nagoya University, Aichi 464-8602, Japan
   \and KMI, Nagoya University, Aichi 464-8602, Japan
   \and Institute of Space and Astronautical Science (ISAS), Japan Aerospace Exploration Agency (JAXA), Kanagawa 252-5210, Japan 
   \and INAF - Istituto di Radioastronomia, via P. Gobetti 101, Bologna, Italy
   \and Department of Theoretical Physics and Astrophysics, Faculty of Science, Masaryk University, Brno 61137, Czechia
   \and Department of Astronomy \& Astrophysics, The University of Chicago, Chicago, IL 60637, USA
   \and Center for Astrophysics Harvard \& Smithsonian, 60 Garden Street,
    Cambridge, MA 02138, USA
    \and Department of Physics, University of Arkansas Main Campus, 825 W Dickson Street Fayetteville, AR 72701, USA
    \and QUP, KEK, 1-1 Oho, Tsukuba, Ibaraki 300-3256, Japan}

   \date{Received September 30, 20XX}

    \abstract
    {The Toothbrush galaxy cluster (1RXS~J0603.3+4214) is known for its unique northern radio relic, central radio halo, and highly disturbed X-ray morphology. It shows features of both shock-driven and turbulent reacceleration, the latter of which is invoked to explain its prominent radio halo. However, it is not yet well understood how the kinetic energy from the merger is distributed into gas motions on various spatial scales.}
   {We aim to probe the dynamics of the intracluster medium (ICM) in the southern, X-ray brightest region of the Toothbrush Cluster by spectrally resolving its bulk and turbulent velocities.} 
   {We analyse a $167$\,ks XRISM/Resolve observation of the cluster in the 3--8\,keV band, and fit the integrated full-array spectrum with single and multi-component thermal models. We also conduct a sub-array analysis by dividing the field of view into quadrants and repeating the fitting procedure while accounting for spatial-spectral mixing.}
   {The full-array spectrum favours a two-component ICM model with a transonic line-of-sight velocity difference of $\Delta v_{\mathrm{bulk}}=1635^{+165}_{-110}$\,$\mathrm{km\,s}^{-1}$, while the one-dimensional line-of-sight velocity dispersions are substantially lower, at $\sigma_{v,1\mathrm{D}}\sim$ 245--310\,$\mathrm{km\,s}^{-1}$. The sub-array analysis likewise reveals a multi-component structure in at least three of the four quadrants, with strong spatial variations in the line-of-sight bulk velocities, reaching $v_{\mathrm{bulk}}=-2520^{+295}_{-230}$\,$\mathrm{km\,s}^{-1}$ to $v_{\mathrm{bulk}}=+930^{+70}_{-110}$\,$\mathrm{km\,s}^{-1}$ relative to the cluster's mean optical redshift ($z_{\mathrm{tooth}} =0. 226$). The velocity dispersions remain modest at 210--215\,$\mathrm{km\,s}^{-1}$, translating to a turbulent pressure support of approximately 3--6\% on scales dominating the observed line broadening. Cooler gas is detected in the NE and SW regions of the pointing, with $kT_{\mathrm{NE}} = 3.13^{+0.33}_{-0.30}$\,keV and $kT_{\mathrm{SW}} = 2.32^{+0.43}_{-0.37}$\,keV respectively; this is likely associated with a third, northern subcluster identified in weak-lensing and optical studies.}
   {Our results indicate a remarkably complex ICM kinematic structure in the Toothbrush Cluster, with evidence of at least three subclusters exhibiting supersonic bulk velocity offsets and significantly smaller velocity dispersions. This shows that the merger geometry has a significant line-of-sight component. It may also indicate that the cluster collision is relatively young, and the turbulent cascade is not yet fully established.}

\titlerunning{ICM Dynamics in the Merging Toothbrush Cluster}

\maketitle
\nolinenumbers

\section{Introduction}

Galaxy clusters are the largest virialized systems that have assembled in the process of cosmic structure formation. They grow through mergers and accretion from the cosmic web \citep{AllenEvrardMantz2011,KravtsovBorgani2012, Voit2005}. The majority of the baryonic content in clusters is dominated by the intracluster medium (ICM), which is a hot diffuse plasma that permeates the space between galaxies \citep{Sarazin1988}. Cluster mergers are the most energetic events to occur in the Universe since the Big Bang and release over $10^{64}$\,erg of gravitational binding energy \citep{Sarazin2002}. This energy can be distributed through several different channels: thermal heating of the ICM/IGM due to shocks and compression \citep{MarkevitchVikhlinin2007, Vazza2017}, turbulent and bulk motions in the ICM on cluster scales \citep{Ryu2008,KravtsovBorgani2012,NormanBryan1999}, amplification of magnetic fields \citep{Roettiger1999}, and acceleration or reacceleration of cosmic-ray particles (\citealt{Ensslin1998,8157925}; for a detailed review, see \citealt{BrunettiJones2014}).

In a merger, infalling substructures initially drive large-scale bulk motions in the gas, which may then feed turbulent cascades \citep{2001ApJ...561..621R,Cassano2005,BrunettiLazarian2007,Vazza2017,Dolag2005,Vazza2009}. However, merger activity in a cluster does not automatically indicate that fully developed turbulence has been established; for example, if the interaction is too recent, the turbulent cascade might not have had sufficient time to transfer kinetic energy to smaller scales. A portion of the energy may also be dissipated at large scales through shocks, viscosity, or other processes. If developed, turbulence can contribute to non-thermal pressure support, mixing metals, regulating transport processes, and accelerating relativistic particles through second-order Fermi processes \citep{Lau2009,BrunettiLazarian2007,BrunettiJones2014,Vazza2018}. 

Diffuse radio emission in galaxy clusters provides a powerful probe of this non-thermal component. On one hand, radio relics are found in the periphery of merging clusters and generally trace shock fronts in which particles are accelerated through the diffusive shock acceleration mechanism \citep{Bell1978}. On the other hand, radio halos roughly trace the ICM baryonic mass distribution and are largely thought to be associated with turbulent reacceleration of relativistic electrons \citep{8157925, Petrosian2001, BrunettiLazarian2007,2015Natur.523...59M,2016MNRAS.458.2584B,vanWeeren2019,2022ApJ...934..182N}. The existence of these sources demonstrates that magnetic fields and cosmic-ray electrons are present on cluster-wide scales. 
Cosmic-ray scattering/transport and turbulent reacceleration mechanisms are dependent on the interaction between particles and magnetic field fluctuations, which in turn are related to velocity perturbations in the ICM. Therefore, combining radio observations with direct ICM velocity measurements provides valuable information for understanding the connection between gas dynamics and particle acceleration. 

Before X-ray microcalorimeters, ICM motions were constrained with very limited spectral precision. XMM-Newton observations of the Coma cluster provided one of the first observational arguments for turbulence in a merging cluster, using spatially resolved pressure maps to infer a fluctuation spectrum that was consistent with a projected Kolmogorov/Oboukhov-like form on scales of tens of kiloparsecs \citep{Schuecker2004}. Subsequent analyses conducted using \textit{Chandra} and XMM-Newton on surface brightness fluctuations converted the 2D fluctuation power spectra into constraints on gas density-perturbations, which in turn could be related to the ICM velocity fluctuations \citep{Zhuravleva2014b,Zhuravleva2015,Zhuravleva2018}. 
However, separating turbulence from sloshing, infalling substructure, abundance variations, and projection effects proved to be difficult \citep{ChurazovEtAl2012, Zhuravleva2023,Heinrich2024}. XMM-Newton/RGS spectroscopic studies placed upper limits on the turbulent velocity broadening in the X-ray emitting gas of relaxed compact cluster cores to a few hundred $\mathrm{km\,s}^{-1}$ \citep{Pinto2015, Pinto2019, SandersEtAl2011}. But since the RGS was slitless, the results were sensitive to spatial broadening caused by the extent of the source along the dispersion direction, rendering measurements for extended, complex merging systems essentially impossible.

High-resolution, non-dispersive X-ray spectroscopy provides the key to probing ICM dynamics. Building on the pioneering measurements of \textit{Hitomi} \citep{Takahashi2014}, XRISM \citep{Tashiro2025} extends direct velocity measurements to a broader sample of galaxy clusters spanning a range of dynamical states. Its Resolve microcalorimeter achieves an energy resolution better than 5\,eV in the 2--12\,keV band \citep{Ishisaki2022}, enabling Doppler shifts and line broadening to be measured with precisions of a few tens of $\mathrm{km\,s}^{-1}$. These observations yield unprecedented constraints on bulk flows and turbulence in galaxy clusters.

Several XRISM observations of merging galaxy clusters have already been carried out. In clusters such as Coma, Abell\,1914, and Abell\,754, the kinetic pressure fraction assuming isotropic turbulence was found to be $\leq$5\% \citep{XRISMComa2025,Heinrich2025,Omiya2026}, which is surprisingly low given the disturbed state of these systems. In the Coma Cluster, XRISM observations revealed line-of-sight (LOS) velocity dispersions of $\sim$200\,$\mathrm{km\,s}^{-1}$, while the bulk velocity offsets spanned several hundred $\mathrm{km\,s}^{-1}$ relative to the optical mean velocity of the cluster. This combination of velocity measurements was initially interpreted as favouring a surprisingly steep turbulent power spectrum compared to the classical Kolmogorov cascade \citep{XRISMComa2025}. However, it was later found that a large turbulent injection scale \citep{Eckert2025} or a spatially-intermittent and multi-scale distribution of turbulence \citep{Vazza2026} can also explain the results while remaining consistent with a Kolmogorov slope. 

In contrast, a velocity dispersion of $\sim$470\,$\mathrm{km\,s}^{-1}$ was found in Abell\,2034, corresponding to a kinetic pressure fraction of $\sim$15\% \citep{2026arXiv260427161H} that is substantially larger compared to other clusters. Therefore, it seems the partition of kinetic energy from the merger between large-scale bulk flows and small-scale isotropic turbulence varies between systems observed by XRISM. These results motivate the need for more ICM velocity measurements, especially in radio-bright mergers where turbulence has been previously invoked to explain particle reacceleration. 

The Toothbrush Cluster, 1RXS J0603.3+4214, at an optical mean redshift of $z_{\mathrm{tooth}} = 0.226$ \citep{2019ApJS..240...39G}, provides us with an exceptional laboratory to study merger-driven ICM dynamics. It hosts one of the most spectacular known radio relics which extends over $1.9$\,Mpc with a nearly linear shape in the northern part of the system, giving the cluster its `Toothbrush' morphology. It also hosts an elongated radio halo and other, smaller relics toward the south and south-east of the cluster \citep{vanWeerenEtAl2012}. Its radio and X-ray morphology indicate that the cluster is in a highly disturbed dynamical state, with previous simulation studies having suggested a complex, and possibly triple merger scenario rather than a binary collision \citep{BruggenvanWeerenRottegering2012}. Previous X-ray studies conducted using XMM-Newton and \textit{Chandra}, as well as weak-lensing studies have shown that the cluster is composed of at least two X-ray-bright cores and mass substructures, and their respective offsets suggest that the system is in a post core-passage merger state \citep{vanWeerenEtAl2016Toothbrush,JeeEtAl2016}.

The Toothbrush Cluster also presents an interesting particle acceleration problem. The radio spectral index at the northern relic suggests a stronger shock than inferred from the X-ray density discontinuity. This produces a tension between the Mach numbers derived from radio and X-ray observations respectively \citep{Ogrean2013,Itahana2015ToothbrushSuzaku,vanWeerenEtAl2016Toothbrush}. However, recent studies have shown that this discrepancy could arise because density inhomogeneities give rise to a distribution of Mach numbers across a shock front, which in turn can result in an over-estimation of the Mach number derived from radio analysis \citep{Whittingham2026}. LOFAR, VLA, and GMRT observations have also shown that the radio spectral index of the cluster flattens to a broad plateau at the location of the radio halo, which suggests that the merger-driven turbulence in the ICM has cascaded to smaller scales and is reaccelerating particles over megaparsec scales \citep{vanWeerenEtAl2012, RajpurohitEtAl2018Toothbrush, deGasperinEtAl2020Toothbrush}. In this work, we use XRISM/Resolve observations of the Toothbrush Cluster to constrain the velocity structure of the ICM. 

This paper is organised as follows: Section\,\ref{section 2} describes the XRISM/Resolve observation and data reduction process, Section\,\ref{section 3} presents the spectral fitting methodology, Section\,\ref{section 4} presents the results from the full- and sub-array analyses, Section\,\ref{section 5} discusses these findings, and lastly, in Section\,\ref{section 6} we summarise our conclusions. 

In this paper, we assume a flat $\Lambda$CDM cosmology, with $H_0 = 70$\,$\mathrm{km\,s}^{-1}\mathrm{Mpc}^{-1}$ and $\Omega_\mathrm{m} = 0.3$, wherein at the redshift of the Toothbrush Cluster ($z_\mathrm{tooth} = 0.226$), 1\,arcmin = 217\,kpc. All reported uncertainties are 1$\sigma$ unless stated otherwise. 


\section{Observations and Data Reduction}\label{section 2}

\begin{figure}
    \centering
    \includegraphics[width=\columnwidth]{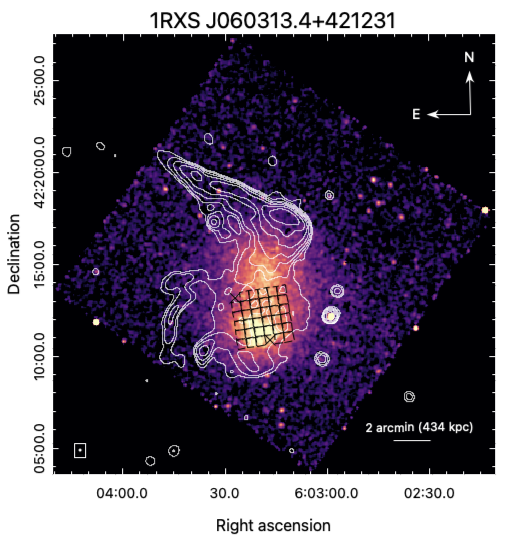}
    \caption{\textit{Chandra} image of the Toothbrush Cluster in the 2--7\,keV band. The XRISM/Resolve field of view is overlaid in black and the individual boxes represent the Resolve instrument's pixelated grid. Pixels 12 and 27 (marked as X's) have been excluded from our analyses. The LOFAR radio contours are overlaid in white at $[1,2,4,8,\ldots]\times4\sigma_{\mathrm{rms}}$ where $\sigma_\mathrm{rms}=0.093$\,mJy\,beam$^{-1}$. The beam size is $8.0^{\prime\prime}\times 6.5^{\prime\prime}$ \citep{vanWeerenEtAl2016Toothbrush}.}
    \label{fov on chandra}
\end{figure}

The Toothbrush Cluster was observed with the XRISM/Resolve instrument between 2025 February 22 and February 25. The Resolve instrument consists of a 6\,$\times$\,6 pixelised X-ray microcalorimeter, with a 3.1$^\prime$\,$\times$\,3.1$^\prime$ field of view, and an average spectral resolution of 4.5\,eV FWHM \citep{2022IJMPD..3130001T}. The observation (\texttt{TOOTHBRUSH\_S}, ObsID: 201032010) was pointed at the southern part of the X-ray-bright central region of the Toothbrush cluster (RA = 90.8305, Dec = 42.2019). The location of this pointing is shown in Fig.\,\ref{fov on chandra}, where it is overlaid (in black) on top of the \textit{Chandra} image of the cluster. 

We analysed the Resolve data using publicly available XRISM analysis tools in HEASoft v6.34. We first reprocessed the data using the \texttt{xapipeline} command (CalDB version 20241115) to apply the latest calibration and standard screening criteria (\texttt{RISE\_TIME, DERIV\_MAX,} and \texttt{STATUS}), following the XRISM data reduction ABC Guide v1.0.\footnote{\href{https://heasarc.gsfc.nasa.gov/docs/xrism/analysis/abc_guide/xrism_abc.html}{XRISM ABC Guide v1.0}} After screening, the net exposure time was 166.7\,ks. We excluded Pixel 27 from all further analysis because it shows anomalous gain variations that cannot be calibrated using the $^{55}$Fe filter wheel data. We also did not consider the data from pixel 12, which is the calibration pixel. The HEASoft tool \texttt{barycen} was used to determine the barycentric correction of $-25.6$\,$\mathrm{km\,s}^{-1}$ for the pointing; this correction has been applied to all bulk velocities that we report in this paper. 

We obtain the spectra using the \texttt{extractor} event extraction tool, selecting only high-resolution primary (or ``Hp'') events. This ensured the best possible spectral resolution while conducting our spectral analysis. We then generate the response matrices using the \texttt{rslmkrmf} task. We use the extra-large \texttt{rmfs} (\texttt{whichrmf} = `\texttt{X}') across our analysis, since this most accurately captures all features of the LSF model, including a Gaussian core, low-energy exponential tail, escape peaks, electron-loss continuum, and Si K fluorescence \citep{Eckart2018, Leutenegger2025}. For XRISM/Resolve data, the \texttt{rmf} input event file is filtered to exclude low-resolution secondary (Ls) events, in order to avoid the effects of false or non-source Ls events and anomalous grade branching ratios.\footnote{NASA, 2026, \textit{HEASoft Help File: rslmkrmf}, \url{https://heasarc.gsfc.nasa.gov/docs/software/lheasoft/help/rslmkrmf.html}.}

We constructed the ancillary response files (ARFs) using the task \texttt{xaarfgen}. This task runs a raytracing simulation using \texttt{xrtraytrace} to calculate how photons from each sky region are distributed across the detector segments due to XRISM's broad point spread function. We use the IMAGE mode of \texttt{xaarfgen}, which requires a surface brightness map of the target as input. For this purpose, we reanalysed the longest existing \textit{Chandra} observation of the Toothbrush Cluster, with ObsID 15172. This data was collected on 2013 November 28 and has an effective exposure time of 111.1\,ks using the ACIS-I chip in very faint (VF) mode. We reprocessed the data using the \texttt{CIAO} software and the dedicated \texttt{chandra\_repro} task with \texttt{check\_vf\_pha=yes}. We then used the \texttt{CIAO} tool \texttt{fluximage} to obtain the flat-fielded surface brightness map of the cluster in the hard band, spanning 2.0--7.0\,keV, roughly matched to XRISM/Resolve's bandpass. This flux map was passed to \texttt{xaarfgen}. 

The non-X-ray background (NXB) was extracted using the \texttt{rslnxbgen} task from the XRISM/Resolve night-Earth database. These event files were also filtered using the same criteria as the observations, and we used v2 of the NXB database.


\section{Spectral fitting}\label{section 3}

We performed spectral fitting in the 3--8\,keV band, using \texttt{XSpec} \citep{Arnaud1996} \texttt{v12.15.0} and adopted the atomic database AtomDB \citep{Foster2012} \texttt{v3.1.2} for plasma modelling. The model fits were evaluated using Cash statistics \citep{1979ApJ...228..939C}. All spectra are optimally binned \citep{KaastraBleeker2016} using the \texttt{ftgrouppha} task in FTOOLS, before any spectral fitting is conducted. 

We model the X-ray emission from the ICM as an optically thin plasma in collisional ionization equilibrium, using one or multiple \texttt{bapec} components, and adopt proto-solar relative abundances with reference to \cite{2009LanB...4B..712L}. 
We model the Galactic absorption using the \texttt{tbabs} model with the hydrogen column density fixed at $3.34 \times 10^{21}$\,atoms\,$\mathrm{cm}^{-2}$ \citep{Willingale2013}. 

The NXB spectrum is fit independently in the 1.7--17\,keV energy range, using a diagonal \texttt{rmf} and a model consisting of a power-law component along with Gaussian profiles for the known detector emission lines. This model is then fixed and included in all spectral fits. The normalizations of the NXB model are scaled according to the number of detector pixels being considered in each (sub-)region. 


\section{Results}\label{section 4}

\subsection{Multi-Velocity Substructure in the Full Field of View}\label{full fov analysis}

\begin{figure*}
    \centering
    \includegraphics[width=0.49\textwidth]{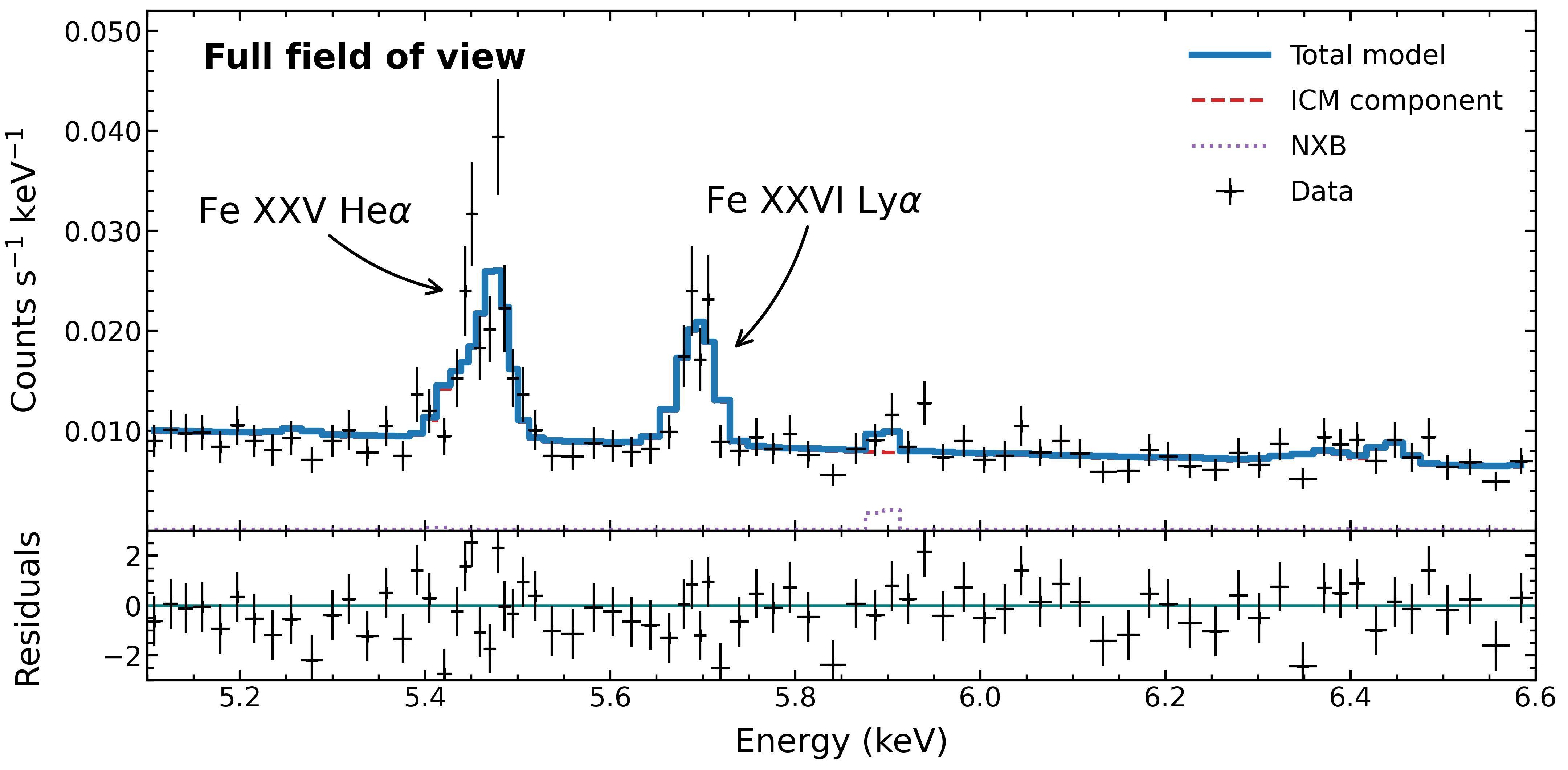}
    \hfill
    \includegraphics[width=0.49\textwidth]{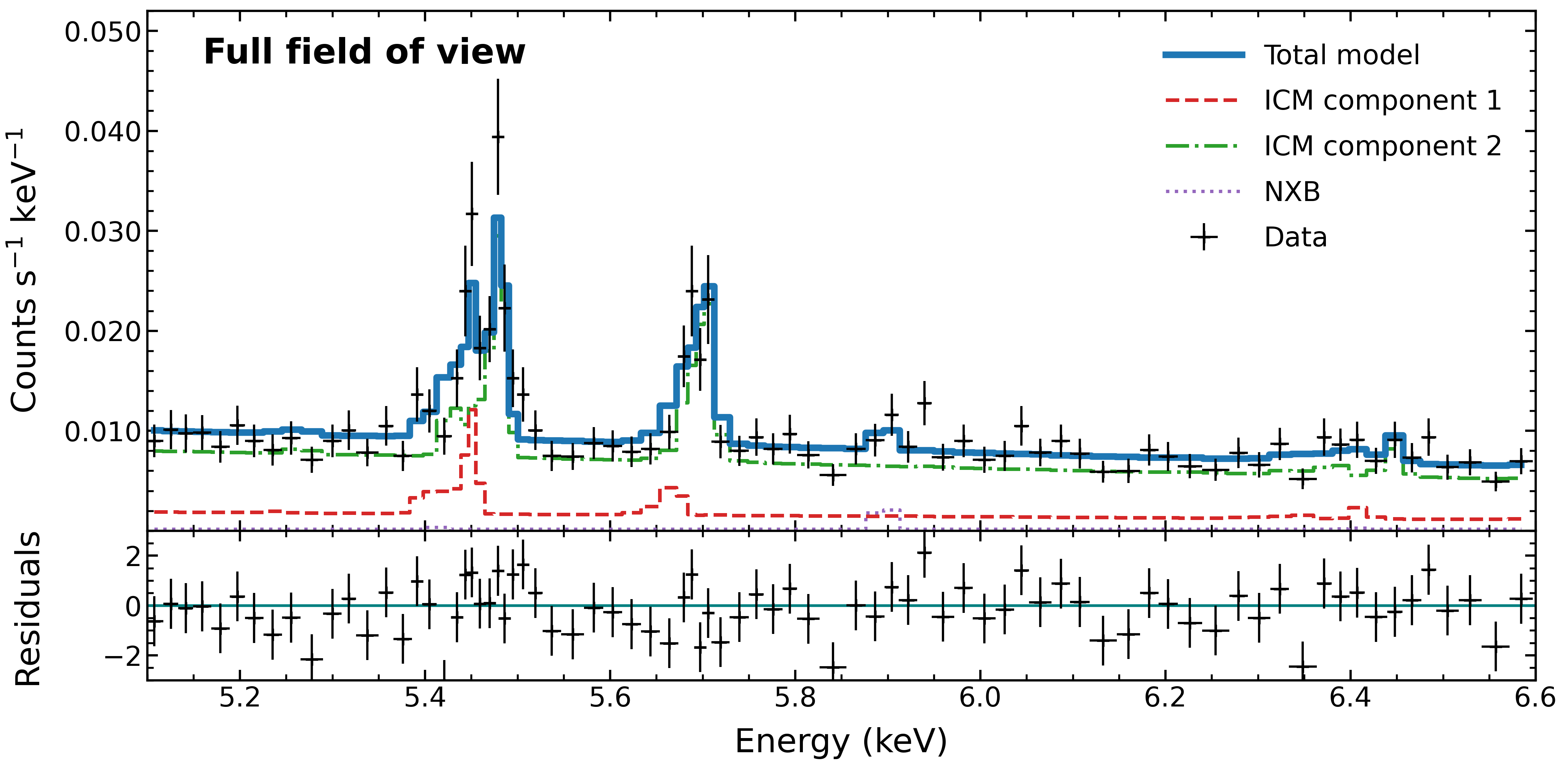}
    \caption{Left: A single \texttt{tbabs$\times$bapec} fit to the full field of view Resolve spectrum of the Toothbrush Cluster, including the NXB model, zoomed into the 5.1--6.6\,keV range. It shows a poor fit to the widths and strengths of the dominant Fe-K lines. Right: A \texttt{tbabs$\times$(bapec+bapec)} fit to the FOV Resolve spectrum of the Toothbrush Cluster, including the NXB model. As seen from the plot, the two-temperature model fits the line strengths and widths in a much better manner than the single-temperature one. The C-stat improves by $\Delta$C-stat = 16.}
    \label{zoomed_fullFOV_1_2_bapec_fits}
\end{figure*}

We begin by fitting the spectrum obtained from the full XRISM/Resolve array. The best-fit model for the full energy band spectra can be seen in the left panel of Fig.\ref{zoomed_fullFOV_1_2_bapec_fits}, where we zoom into the Fe-K complex (5.1--6.6\,keV). For the full-band spectra, we refer the reader to Appendix\,\ref{full fov full band spectra}. At first, we model the spectrum assuming a single ICM component plus the NXB. However, the single-phase model does not describe the spectral features accurately, as the line widths and strengths are fit poorly, especially for the Fe\,XXV\,He$\alpha$ and Fe\,XXVI\,Ly$\alpha$ lines.

Considering the residuals around these lines, we fit the spectrum with a two-component ICM model (\texttt{tbabs$\times$(bapec+bapec)}), which improves the fit by $\Delta$C-stat\,$=$\,16. This model is plotted in the right panel of Fig.\ref{zoomed_fullFOV_1_2_bapec_fits} as a blue line, with its two velocity components represented by the red and green dashed lines respectively. To better gauge which of these models is favourable, we also compute the Akaike Information Criterion (AIC) for each, which is given by \citep{1979ApJ...228..939C, 2017A&A...605A..51K},
\begin{equation}
    \mathrm{AIC} = 2k + C,
\end{equation}
where $k$ is the number of parameters and $C$ is the C-stat. The first single-component model has $k=5$ and C-stat = 1354, while the two-component model has $k = 9$ and C-stat = 1338. This implies that for the first model, we get $\mathrm{AIC}_{1C} = 1364$ and for the second $\mathrm{AIC}_{2C} = 1356$. Therefore, $\Delta\mathrm{AIC} = 8$ indicates that the second, two-component model, which has a lower AIC, is meaningfully better than the single-component model. 

The best-fit parameters for both models have been reported in Table\,\ref{full field of best fit parameters for both models}. We observe two line-of-sight bulk velocity components, at redshifts $z_1=0.2293^{+0.0006}_{-0.0004}$ and $z_2=0.2226^{+0.0003}_{-0.0002}$ respectively. This means that the first component is redshifted by $v_{\mathrm{bulk},1}=780^{+145}_{-100}$\,$\mathrm{km\,s}^{-1}$ while the second component is blueshifted by $v_{\mathrm{bulk},2}=-855^{+75}_{-50}$\,$\mathrm{km\,s}^{-1}$ relative to the optical mean redshift of the cluster. This implies a line-of-sight velocity separation of $\Delta v_{\mathrm{bulk}}=1635^{+165}_{-110}$\,$\mathrm{km\,s}^{-1}$ between the two components. 
Despite this transonic line-of-sight velocity difference, their velocity dispersions remain substantially lower, with the first component showing a dispersion of $\sigma_v = 245^{+105}_{-175}$\,$\mathrm{km\,s}^{-1}$, while the velocity dispersion of the second is $\sigma_v=310^{+100}_{-60}$\,$\mathrm{km\,s}^{-1}$. Contrarily, in the case of the single-component model, we observe an inflated velocity dispersion of $\sigma_v=750^{+165}_{-375}$\,$\mathrm{km\,s}^{-1}$, 
likely driven by the superposition of motions along the line-of-sight.  
The best-fit metallicity corresponding to this model is $0.28^{+0.03}_{-0.03}\,Z_\odot$, coupled between the two components to ensure fit stability. This effectively represents the Fe/H abundance, since iron has the strongest emission lines in the observed spectra. This abundance value is in good agreement with previous X-ray studies conducted using \textit{Chandra} and XMM-Newton \citep{vanWeerenEtAl2016Toothbrush, Ogrean2013, Urdampilleta2019}.

\begin{table*}
\centering
\caption{Best-fit parameters for the one- and two-component ICM model ((a) \texttt{tbabs$\times$bapec} and (b) \texttt{tbabs$\times$(bapec+bapec)}), for the full field of view configuration of the Toothbrush Cluster. The line-of-sight bulk velocity was computed from the redshift using the formula $v_{\rm bulk} = (z-z_\mathrm{tooth})\times c/(1+z_\mathrm{tooth})$ where $z_\mathrm{tooth}=0.226$.}
\label{full field of best fit parameters for both models}
\renewcommand{\arraystretch}{1.6}
\resizebox{\textwidth}{!}{%
\begin{tabular}{lccccccc}
\hline
\hline
\textbf{\rule{0pt}{4ex}} 
& \shortstack{\rule{0pt}{4ex}\textbf{Temperature}\\\textbf{(keV)}} 
& \textbf{\rule{0pt}{4ex}Abundance} 
& \textbf{\rule{0pt}{4ex}Redshift} 
& \shortstack{\rule{0pt}{4ex}\textbf{Bulk Velocity}\\\textbf{($\mathrm{km\,s}^{-1}$)}} 
& \shortstack{\rule{0pt}{4ex}\textbf{$\sigma_v$}\\\textbf{($\mathrm{km\,s}^{-1}$)}} 
& \textbf{\rule{0pt}{4ex}Normalisation} 
& \shortstack{\rule{0pt}{4ex}\textbf{C-stat}\\\textbf{(Expected C-stat)}} \\
\hline
\textbf{1 Component Model} 
& $8.51_{-0.38}^{+0.41}$ 
& $0.29_{-0.05}^{+0.04}$ 
& $0.2233_{-0.0005}^{+0.0004}$ 
& $-685_{-120}^{+100}$ 
& $750_{-375}^{+165}$ 
& $3.27_{-0.09}^{+0.09}\times10^{-2}$ 
& $1354~(1296 \pm 52)$ \\
\hline
\textbf{2 Component Model} & & & & & & & \\
Component 1 
& $6.78_{-1.68}^{+1.65}$ 
& $0.28_{-0.03}^{+0.03}$ 
& $0.2293_{-0.0004}^{+0.0006}$ 
& $780_{-100}^{+145}$ 
& $245_{-175}^{+105}$ 
& $7.23_{-2.08}^{+1.88}\times10^{-3}$ 
& $1338~(1296 \pm 52)$ \\
Component 2 
& $9.09_{-0.68}^{+0.73}$ 
& $0.28$ (coupled) 
& $0.2226_{-0.0002}^{+0.0003}$ 
& $-855_{-50}^{+75}$ 
& $310_{-60}^{+100}$ 
& $2.55_{-0.19}^{+0.22}\times10^{-2}$ 
& \\
\hline
\end{tabular}%
}
\end{table*}

We note that the temperatures of the two components are separated at only $1.29\sigma$. A fit performed while coupling the temperatures across the two components yields similar values of the redshifts, abundances, norms, and dispersions as the original fit, and there is no significant change in the C-statistics (see Appendix\,\ref{fullarray temp tied} for the best-fit model). Therefore, at this point the data do not seem to indicate the need for a two-component model due to a statistically significant temperature contrast, but rather to account for the kinematic separation of the two components.

\subsection{Spatial Distribution of ICM Velocities} \label{spatial dist of vel}

Although the full field of view analysis reveals two velocity substructures within the ICM, it does not inform us about their spatial distribution across the pointing. We therefore divide the Resolve field of view into quadrant subregions to gain a more spatially resolved picture of the ICM dynamics and thermal structure within the Toothbrush Cluster. The four quadrants are named according to their cardinal directions in the sky: south-east (SE), north-east (NE), north-west (NW), and south-west (SW). 

\begin{figure}
    \centering
    \includegraphics[width=\columnwidth]{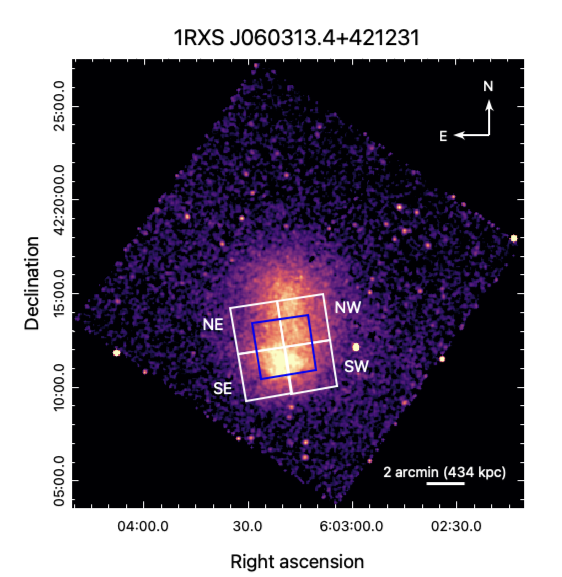}
    \caption{\textit{Chandra} image of the Toothbrush cluster in the 2--7\,keV band. The XRISM/Resolve field of view is overlaid in purple, while the sky regions are shown in white. As discussed previously, pixels 12 and 27 have been excluded from our analysis.}
    \label{Tooth_subregions}
\end{figure}

We fit the spectra from the four sub-array quadrants simultaneously. To model spatial-spectral mixing (SSM), the corresponding sky regions are chosen as shown in Fig.\,\ref{Tooth_subregions}. The total sky region is a box with a side length of 5\,arcmin, centred around and projected onto the same sky coordinates as the XRISM/Resolve field of view. Each sky subregion is chosen to closely trace the corresponding detector region along two edges, while slightly extending beyond it along the outer edges. This allows us to account for scattering from regions just beyond the XRISM/Resolve field of view as well. 

As an initial diagnostic, each quadrant is first fit with a single temperature \texttt{crossarf$\times$tbabs$\times$bapec} model to assess the broad distribution of ICM properties (see Appendix\,\ref{single temp model quadrants} for this model). This best-fit reveals large apparent velocity dispersions in the NE ($\sim$360\,$\mathrm{km\,s}^{-1}$) and NW ($ >$600\,$\mathrm{km\,s}^{-1}$) regions. Rather than attributing these large dispersions to unusually strong isotropic turbulence, we check whether they arise from velocity substructure along the line-of-sight, based on our integrated full-array analysis. 

To this end, we then fit the spectra with a two component \texttt{crossarf$\times$tbabs(bapec+bapec)} model across the four regions. Initially, only the component normalizations, redshifts, and temperatures are allowed to vary freely. We tie the velocity dispersions and abundances of each component across the four quadrants to ensure the convergence of the fit. In the NE, NW, and SW regions, both components are significantly detected and show distinct LOS kinematic properties. In the SE region, the additional component has a best-fit normalization that is about an order of magnitude smaller than the main component. Therefore, we do not consider the second component to be statistically required in this region and we consequently remove it from further modelling. The more strongly detected component does not show significant temperature variation across the field of view and the temperatures measured within the individual regions are mutually consistent within their 1$\sigma$ uncertainties (see Appendix\,\ref{before tying temepratures of 3+4 model}). We therefore tie the temperature of this component across all regions. The temperatures of the second component are left to vary freely due to the detection of significant thermal structure. The best-fit parameters from this model have been reported in Table\,\ref{quadrant_best_fit_new_with_region_kt}. Hereafter, we refer to this as `model A'.

\begin{table}[!ht]
\caption{Best-fit parameters for the quadrant sub-array analysis of the Toothbrush Cluster using the two-temperature \texttt{crossarf$\times$tbabs(bapec+bapec)} model (`model A') in the full 3--8\,keV band. The line-of-sight bulk velocities are computed from the redshift using the formula $v_{\rm bulk}=(z-z_\mathrm{tooth})\times c/(1+z_\mathrm{tooth})$.}
\label{quadrant_best_fit_new_with_region_kt}

\centering
\renewcommand{\arraystretch}{1.4}

\begin{tabular}{lcc}
\hline\hline
Parameter (Unit)
& \shortstack{Main\\component}
& \shortstack{Secondary\\component} \\
\hline

$kT$ (keV) 
& $9.50^{+0.35}_{-0.36}$ 
& -- \\

Abundance ($Z/Z_\odot$) 
& $0.31^{+0.03}_{-0.03}$ 
& $0.40^{+0.06}_{-0.07}$ \\

$\sigma_v$ ($\mathrm{km\,s}^{-1}$) 
& $210^{+30}_{-75}$ 
& $215^{+50}_{-95}$ \\
\hline

\multicolumn{3}{l}{\textbf{SE Region}} \\

$kT$ (keV) 
& $9.50$ (coupled) 
& -- \\

Redshift 
& $0.2224^{+0.0002}_{-0.0005}$ 
& -- \\

$v_{\rm bulk}$ ($\mathrm{km\,s}^{-1}$) 
& $-905^{+50}_{-130}$ 
& -- \\

Norm ($10^{-3}$) 
& $2.32^{+0.07}_{-0.08}$ 
& -- \\
\hline

\multicolumn{3}{l}{\textbf{NE Region}} \\

$kT$ (keV) 
& $9.50$ (coupled) 
& $3.13^{+0.33}_{-0.30}$ \\

Redshift 
& $0.2229^{+0.0005}_{-0.0010}$ 
& $0.2173^{+0.0006}_{-0.0006}$ \\

$v_{\rm bulk}$ ($\mathrm{km\,s}^{-1}$) 
& $-785^{+135}_{-240}$ 
& $-2155^{+150}_{-160}$ \\

Norm ($10^{-3}$) 
& $1.16^{+0.10}_{-0.09}$ 
& $1.499^{+0.212}_{-0.236}$ \\
\hline

\multicolumn{3}{l}{\textbf{NW Region}} \\

$kT$ (keV) 
& $9.50$ (coupled) 
& $8.06^{+0.89}_{-0.88}$ \\

Redshift 
& $0.2253^{+0.0003}_{-0.0009}$ 
& $0.2299^{+0.0003}_{-0.0005}$ \\

$v_{\rm bulk}$ ($\mathrm{km\,s}^{-1}$) 
& $-195^{+80}_{-225}$ 
& $930^{+70}_{-110}$ \\

Norm ($10^{-3}$) 
& $1.15^{+0.11}_{-0.09}$ 
& $1.501^{+0.100}_{-0.300}$ \\
\hline

\multicolumn{3}{l}{\textbf{SW Region}} \\

$kT$ (keV) 
& $9.50$ (coupled) 
& $2.32^{+0.43}_{-0.37}$ \\

Redshift 
& $0.2228^{+0.0002}_{-0.0007}$ 
& $0.2158^{+0.0012}_{-0.0009}$ \\

$v_{\rm bulk}$ ($\mathrm{km\,s}^{-1}$) 
& $-810^{+50}_{-165}$ 
& $-2520^{+295}_{-230}$ \\

Norm ($10^{-3}$) 
& $1.92^{+0.09}_{-0.08}$ 
& $0.943^{+0.335}_{-0.274}$ \\
\hline

C-stat (Expected C-stat) 
& \multicolumn{2}{c}{$4716$ ($4762 \pm 91$)} \\

DOF 
& \multicolumn{2}{c}{4168} \\
\hline

\end{tabular}
\end{table}

In our analysis, the main component has a best-fit temperature of $kT_\mathrm{main}=9.50^{+0.35}_{-0.36}$\,keV which is in good agreement with previous X-ray analysis results using XMM-Newton, \textit{Suzaku}, and \textit{Chandra}, and they all consistently found temperatures of $\sim$7--10\,keV in the regions covered by our XRISM/Resolve pointing \citep{Ogrean2013, Itahana2015ToothbrushSuzaku, vanWeerenEtAl2016Toothbrush}.   
On the other hand, the secondary component shows significant spatial variation in terms of its thermal structure: in the NE and SW regions, it is significantly cooler than the cluster average at $kT_\mathrm{second,NE}=3.13^{+0.33}_{-0.30}$\,keV and $kT_\mathrm{second,SW}=2.32^{+0.43}_{-0.37}$\,keV, while in the NW region it is hot and comparable to the temperature of the main component at $kT_\mathrm{second,NW}=8.06^{+0.89}_{-0.88}$\,keV. In the diagnostic single-component modelling that was conducted (see Appendix\,\ref{single temp model quadrants}), the NE region already showed a cooler than average temperature of $kT\sim$5.17\,keV. Introducing the second component in model A simply separated the two phases further.

\begin{figure*}
    \centering
    \makebox[0.8\textwidth][c]{%
        \includegraphics[width=0.95\textwidth]{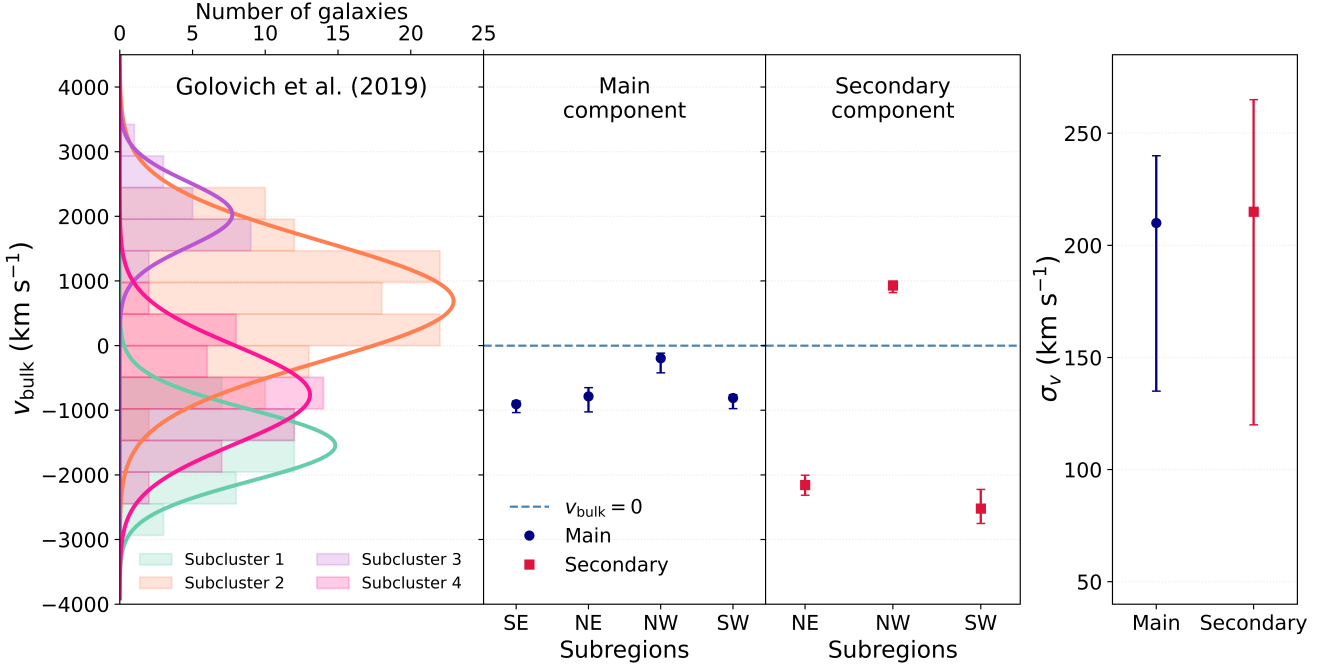}}
    \caption{The plots present a comparison between the bulk velocities and velocity dispersions of the main and secondary components from model A with the optical galaxy redshift measurements. The leftmost panel shows the spectroscopic velocity distributions of 242 galaxies divided amongst four subclusters (as detailed in \citealt{2019ApJ...882...69G}). The main X-ray component appears consistent with Subcluster 4, while the secondary component in the NE and SW regions are consistent with Subcluster 1, and the redshifted phase in the NW is consistent with Subcluster 2. Despite the supersonic velocity offsets observed for both components, their dispersions remain low/moderate at $\sim$215\,$\mathrm{km\,s}^{-1}$. A deeper discussion of this comparison is presented in Section\,\ref{section 5.1}.}
    \label{bulk_and_turb_velocities_depiction}
\end{figure*}

The ICM components in the Toothbrush Cluster also show strong spatial substructure in terms of the line-of-sight bulk velocities across the four regions. The main component is blueshifted across all regions, with bulk velocities reaching $v_{\mathrm{bulk,SE}}=-905^{+50}_{-130}$\,$\mathrm{km\,s}^{-1}$, $v_{\mathrm{bulk,NE}}=-785^{+135}_{-240}$\,$\mathrm{km\,s}^{-1}$, $v_{\mathrm{bulk,NW}}=-195^{+80}_{-225}$\,$\mathrm{km\,s}^{-1}$, and $v_{\mathrm{bulk,SW}}=-810^{+50}_{-165}$\,$\mathrm{km\,s}^{-1}$ respectively. The secondary component shows even larger variations. While the gas is strongly blueshifted in the NE and SW regions, at $v_\mathrm{bulk} = -2155^{+150}_{-160}$\,$\mathrm{km\,s}^{-1}$ and $v_\mathrm{bulk} = -2520^{+295}_{-230}$\,$\mathrm{km\,s}^{-1}$, it exhibits a strong redshift in the NW region at $v_\mathrm{bulk} = 930^{+70}_{-110}$\,$\mathrm{km\,s}^{-1}$. The dynamics of gas in the NW region is rather interesting, as it suggests there is some bulk motion that causes both component gas phases to shift towards higher redshift velocities when compared with other regions. The bulk and turbulent velocities presented in Table\,\ref{quadrant_best_fit_new_with_region_kt} for the fiducial model\,A have been plotted in Fig.\,\ref{bulk_and_turb_velocities_depiction} for the main and secondary ICM components for ease of comparison.

Due to the calibration uncertainties in the XRISM X-ray Mirror Assembly and its associated point spread function, we test the robustness of the SSM analysis by scaling the cross-ARFs by $\pm$30\% (summed over pixels, as previously done in e.g. \citealt{XRISM2026Perseus, XRISM2025A2029}). We modify the relative contribution of photons scattered from adjacent regions and repeat the spectral fitting procedure. Doing so, it was found that all parameters from model A remain consistent with each other within their 1$\sigma$ statistical uncertainties (see Appendix \ref{appendix sys unc data}). Therefore, the conclusion of our study is robust to these systematic uncertainties.

\subsection{Comparison of image-based and spectral-based component separation methods}\label{checking if vel of beta model are robust}

The multiple velocity components reported above suggest that we may be seeing kinematic signatures of (at least) the two main subclusters of the Toothbrush merging system, oriented along the north-south axis. To confirm if this interpretation is indeed consistent with the X-ray morphology of the Toothbrush Cluster, we attempt to disentangle the two components also in image space, and compare the flux ratios to the spectral fit results. 

\begin{figure*}[!h]
    \centering
    \includegraphics[width=0.33\textwidth]{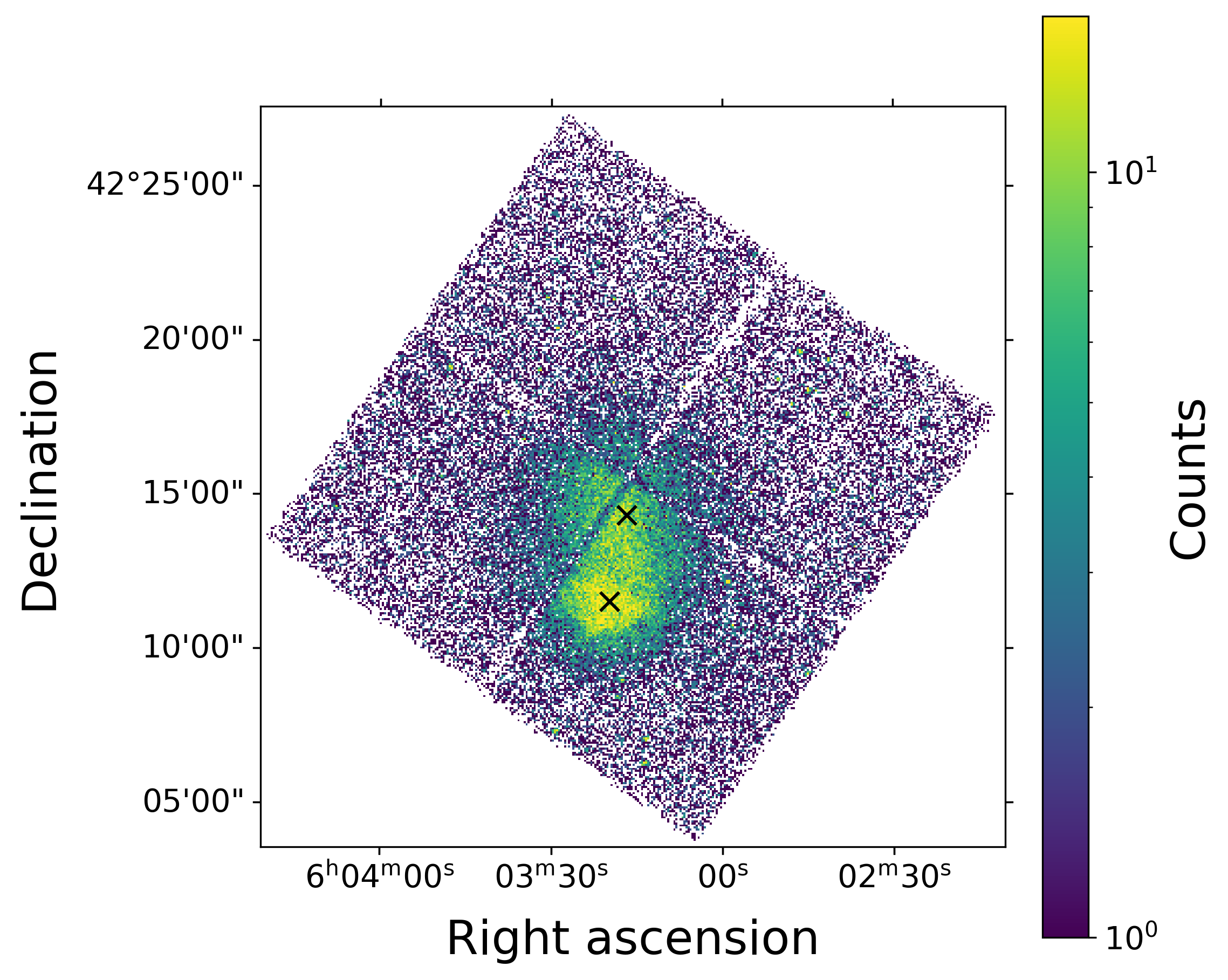}
    \hfill
    \includegraphics[width=0.33\textwidth]{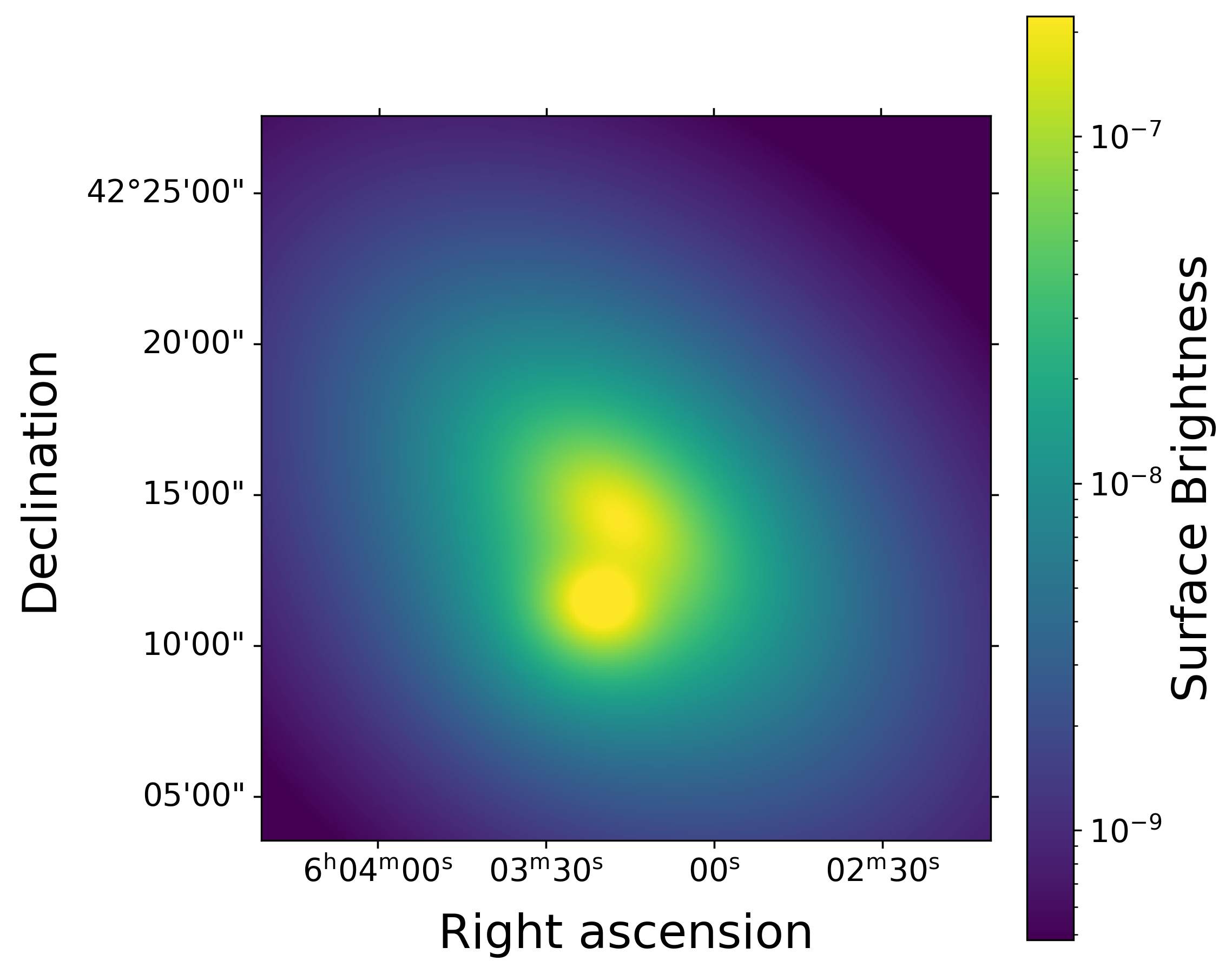}
    \hfill
    \includegraphics[width=0.33\textwidth]{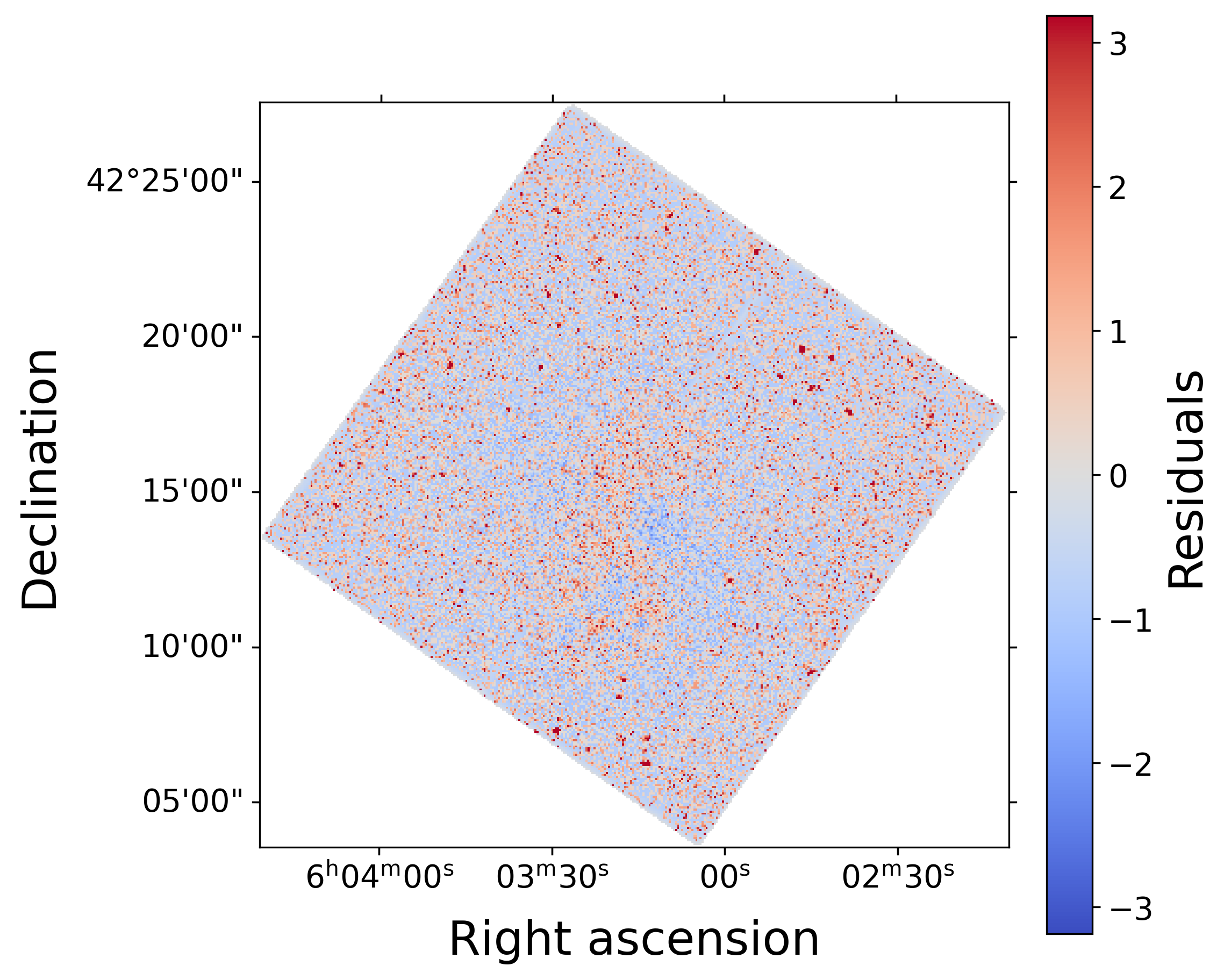}
    \caption{Left: Observed \textit{Chandra} count map, with the black crosses marking the positions of the two X-ray peaks. Middle: Best-fit double $\beta$-model with different peak coordinates, core radii, power-law slopes, and ellipticities. Right: Residual map given by \textit{Sherpa} for the double $\beta$-model fit. We do not observe any significant residual structures.}
    \label{fig:three_images}
\end{figure*}

We fit the \textit{Chandra} hard-band counts map with two elliptical $\beta$-model components that have different centres (position of the peaks), power-law slopes, ellipticities, position angles, and core radii, plus a uniform background using the \textit{Sherpa} software \citep{2024ApJS..274...43S, 2009SciPy...51R, 2001SPIE.4477...76F} in \texttt{CIAO} \texttt{v4.17} \citep{Fruscione2006} and the calibration data \texttt{CALDB} \texttt{v4.12} \citep{2006SPIE.6270E..1XG}. The model is also multiplied by the \textit{Chandra} exposure map of the cluster and is given as, 
\begin{equation}
    M(x,y) = E(x,y)\,\Big(S_N(x,y)+S_S(x,y) +B\Big),
\end{equation}
where, $E(x,y)$ is the exposure map, $B$ is the uniform background, and $S_N$ and $S_S$ are the $\beta$-model components associated with the northern and southern subclusters, respectively. The functional form of the surface brightness profile of each component fitted using the \textit{Sherpa} \texttt{beta2d} model is, 
\begin{equation}
    S_i(x,y) = A_i\left(1+\frac{r_i^2}{r_0^2}\right)^{-\alpha_i},
\end{equation}
where $A_i$ is the model value at the peak position, $\alpha_i$ is the power law slope at large-radii, $r_i$ is the model radius, and $r_0$ is the core radius. The fit is evaluated using Cash statistics, and we allow the peak coordinates to vary freely. Upon doing so, we find that both peaks coincide with the two main X-ray cores of the Toothbrush Cluster that survived the merger. These have been marked by the black crosses in Fig.\,\ref{fig:three_images}. We also report the best-fit parameters of the double $\beta$-model in Table\,\ref{tab:double_beta_2d_fit}. We do not observe any significant residuals in the residual map, suggesting that the double $\beta$-model is an adequate representation of the cluster substructures.

\begin{table*}
\centering
\caption{Best-fit parameters of the double-$\beta$ model fit to the \textit{Chandra} image of the Toothbrush Cluster in the 2--7\,keV band. The model used is \texttt{emap$\times$($\beta_1$+$\beta_2$+Bg)}. The errors reported correspond to the 1$\sigma$ confidence range in \textit{Sherpa}. The amplitude $A$ is reported in units of $[10^{-7}\mathrm{~photons}\mathrm{~cm}^{-2}\mathrm{~pixel}^{-2}\mathrm{~s}^{-1}]$.}
\label{tab:double_beta_2d_fit}
\renewcommand{\arraystretch}{1.35}
\begin{tabular}{lccccc}
\hline\hline
Component & $\alpha$ & $r_0$ (kpc) & Amplitude & Ellipticity & $\theta$ (rad)
\\
\hline
$\beta_1 ~(\mathrm{Southern ~Subcluster})$ & $1.80 \pm 0.06$ & $292.6 \pm 8.7$ & $4.16 \pm 0.08$ & $0.09 \pm 0.02$ & $-0.20 \pm 0.08$
\\
$\beta_2 ~(\mathrm{Northern~Subcluster})$
& $1.44 \pm 0.03$
& $520.9 \pm 12.5$
& $2.11 \pm 0.04$
& $0.33 \pm 0.01$
& $-0.73 \pm 0.02$
\\
\hline
\end{tabular}
\end{table*}

Over the full XRISM/Resolve field of view, the double $\beta$-model predicts that the southern component ($\beta_1$) has a fractional contribution of 0.72, while that of the northern component ($\beta_2$) is 0.28. This implies a corresponding ratio of $\beta_2/\beta_1 = 0.38$.
We compare this with the two-temperature full-array spectral fitting results reported in Table\,\ref{full field of best fit parameters for both models}, from which we get $\mathrm{N}_1/\mathrm{N}_2 = 0.28^{+0.08}_{-0.09}$. Therefore, the relative contribution of the two components inferred from the \textit{Chandra} image is broadly consistent with that obtained from the spectral decomposition across the full FoV. 

We then use the double-$\beta$ model to estimate the relative contributions of each component within the sky regions (SW, SE, NW, and NE) shown in Fig. \ref{Tooth_subregions}. This is done using the \texttt{dmstat} tool in \texttt{CIAO}. We sum the emission from each $\beta$-model component within the four quadrants and report their respective contribution ratios in Table \ref{double beta contributions}. This table also reports the relative contributions of each velocity component, based on their respective normalisations across quadrants, from our fiducial `model A'.

\begin{table}
\caption{The ratio of the relative contributions of the southern ($\beta_1$) and northern subclusters ($\beta_2$) in each quadrant defined in Fig.\,\ref{Tooth_subregions}, as estimated using the double-$\beta$ surface-brightness model. 
The rightmost column gives the corresponding ratios of the best-fit normalisations of the main and secondary 
\texttt{bapec} components in each region from model A reported in Table\,\ref{quadrant_best_fit_new_with_region_kt}. }
\label{double beta contributions}

\centering
\renewcommand{\arraystretch}{1.3}
\setlength{\tabcolsep}{6pt}

\begin{tabular}{lcc}
\hline\hline
\noalign{\vskip 2pt}
Region
& \shortstack{Surface Brightness\\ Ratio ($\beta_{1}/\beta_{2}$)}
& \shortstack{Spectral ratio from model A\\($\mathrm{Norm}_{\mathrm{main}}/\mathrm{Norm}_{\mathrm{second}}$)} \\
\hline
SE & 8.34 & -- \\
NE & 0.91 & $0.77^{+0.14}_{-0.12}$ \\
NW & 0.22 & $0.76^{+0.17}_{-0.08}$ \\
SW & 2.33 & $2.04^{+0.60}_{-0.73}$ \\
\hline
\end{tabular}
\label{ratio of component contrubutions}
\end{table}

Based on the results reported in Table\,\ref{ratio of component contrubutions}, in the NE region, the predicted $\beta$-model ratio of 0.91 is consistent with the spectral best-fit value of $0.77^{+0.14}_{-0.12}$. Similarly, in the SW region, the predicted ratio of 2.33 agrees with the best-fit value of $2.04^{+0.60}_{-0.73}$ within its statistical uncertainties. In the SE region, the double $\beta$-model predicts that the northern component only contributes to about 11\% of the total emission. Consistent with this expectation, in model A, the second spectral component was too weak to be constrained while conducting the spectral fitting and was therefore excluded. However, the NW region shows a clear discrepancy. While the double $\beta$-model predicts a southern-to-northern component ratio of only $0.22$, the model A spectral normalisation ratio is $0.76^{+0.17}_{-0.08}$. However, it is important to note that the double $\beta$-model provides only an approximate description of the complex gas morphology within a highly disturbed merging cluster, likely hosting more than two subhalos (see Section \ref{section 5.1}). Additionally, it is possible that a part of this discrepancy arises because the spectral decomposition into a two-temperature model is in itself uncertain. Therefore, despite the mismatch in the NW region, we can conclude that the relative contributions of the two substructures inferred from the $\beta$-model analysis and model\,A broadly agree. 

To test the robustness of our results, we repeated the spectral analysis with component normalisations tied based on the flux ratios predicted by the double $\beta$-model (hereafter `model B', see Appendix \ref{model B}). Although this approach produced a significantly worse fit ($\Delta$C-stat = +16) than our fiducial `model A', the inferred bulk velocities and velocity dispersions remained consistent within their $1\sigma$ uncertainties across all sub-regions (see Fig. \ref{before and after temp are tied vel map}). Importantly, this agreement persists despite `model B' imposing a substantially different temperature structure, indicating that the observed velocity pattern is insensitive to both the adopted component separation and potential temperature-redshift degeneracies. We therefore conclude that the measured velocity structure is robust, while retaining `model A' as our preferred description of the data.


\section{Discussion}\label{section 5}

\subsection{Merger Geometry}\label{section 5.1}

Previous XMM-Newton and \textit{Chandra} studies describe the Toothbrush Cluster as a disturbed system, with two major surviving ICM cores, and a bullet-like southern X-ray structure, bounded by a cold front \citep{vanWeerenEtAl2016Toothbrush, Ogrean2013}. Weak lensing analysis showed that the dominant mass distribution is bimodal along the north-south axis, with an offset between the mass and X-ray peaks: the northern mass peak is offset towards the northwest with respect to the corresponding X-ray peak by $\sim$\,0.5$^\prime$, whereas the southern mass peak is offset to the south by $\sim$\,2$^\prime$ \citep{JeeEtAl2016}. This offset has been interpreted as evidence for a post-core passage cluster merger in which the two components have passed through each other and are still separating. Early optical redshift analyses had revealed a significant line-of-sight velocity difference, of $\sim$1800\,$\mathrm{km\,s}^{-1}$, between the two main subclusters \citep{JeeEtAl2016}. 

Our XRISM integrated spectrum also reveals two components with a large relative line-of-sight velocity difference of 1635\,$\mathrm{km\,s}^{-1}$, where the cooler ICM component is redshifted by 780\,$\mathrm{km\,s}^{-1}$, while the hotter component is blueshifted by 855\,$\mathrm{km\,s}^{-1}$ with respect to the cluster's mean optical rest-frame. The strong agreement between the velocity difference inferred from our XRISM analysis and that from galaxy redshifts seemingly suggests that the cluster merger has a significant line-of-sight component. However, this picture is in tension with previous radio studies. It was believed that the merger is happening very nearly in the plane of the sky as the frontal part of the main relic is highly polarized (up to $\sim$60\%; \citealt{vanWeerenEtAl2012}). Therefore, while the high radio polarization supports a merger geometry close to the plane of the sky, the X-ray and optical analyses indicate significant line-of-sight  components.

This LOS inclination has important implications also for estimating the age of the merger. Hydrodynamical simulations conducted in \citet{BruggenvanWeerenRottegering2012} and assuming a plane of the sky geometry suggested that core passage in the Toothbrush Cluster occurred roughly 2 Gyr ago, implying a system currently approaching virialization. Instead, assuming an angle of $\sim30^\circ$ between the line of sight and the normal to the merger plane, and using the measured relative velocity of 1635 km s$^{-1}$, and a projected separation of 650 kpc \citep{Ogrean2013}, we now estimate a time since core passage of only $\sim0.2$ Gyr.

\subsubsection*{Origin of the Velocity Structures}\label{origin of velocity compoennts}

The velocity decomposition obtained from the quadrant analysis (model A) suggests that the two spectral components are associated with the cluster's two main subclusters. The primary component is detected across all four quadrants, whereas the secondary component is significantly detected only in the NE, NW, and SW regions, which lie toward the northern half of the XRISM field of view. Given that the Resolve pointing is centred on the southern X-ray-bright region, with the southern X-ray core located mainly in the SE and SW quadrants, the primary component is naturally associated with the southern subcluster. The secondary component can then be interpreted as gas belonging to the northern subcluster projected along our line of sight. 

This hypothesis is further strengthened by the double $\beta$-model image analysis we conducted. The two components of the best-fit $\beta$-model are centred on the northern and southern X-ray cores and their relative flux contributions broadly agree with the normalisation ratios obtained using the spectral decomposition in model A across most of the Resolve field of view. Therefore, the two velocity components detected by XRISM are most naturally interpreted as emission from the two main subclusters, that underwent collision along the north-south axis. 

To gain further insight into the merger substructure, we compare our results with the optical galaxy kinematics presented by \citet{2019ApJ...882...69G}. Using 242 galaxy redshifts, that study identified four subclusters associated with the mass concentrations detected in the weak-lensing analysis of \citet{JeeEtAl2016}: two structures in the north (Subclusters 1 and 2), a dominant southern structure (Subcluster 4), and a less well-constrained component west of the radio halo (Subcluster 3).

A striking correspondence emerges between the optical and X-ray velocity distributions. As shown in Fig.~\ref{bulk_and_turb_velocities_depiction}, the primary XRISM component has redshifts consistent with those of optical Subcluster 4, reinforcing its identification with the southern subcluster. The secondary component exhibits a more complex behaviour. Its blueshifted phases in the NE and SW quadrants are consistent with Subcluster 1, whereas its redshifted phase in the NW quadrant is consistent with Subcluster 2. Since both optical subclusters are located in the northern part of the system, the XRISM "secondary" component may in fact represent a superposition of multiple northern merger constituents rather than a single ICM structure. However, retracing back to our $\beta$-model analysis, these three components can not be clearly separated in the \textit{Chandra} surface brightness image (see Fig.\,\ref{fov on chandra}), since we observe only two distinct X-ray peaks. But we note that since the additional northern subcluster lies very close to the main northern peak, it is possible that it may be contributing to the significant ellipticity/elongation of the northern $\beta$-component (see Table\,\ref{tab:double_beta_2d_fit}), instead of appearing as a third separate X-ray peak.  

Such an interpretation could explain several features of the data. The cooler gas detected in the NE and SW quadrants may be associated with the optically poorer, and likely less massive, Subcluster 1. The discrepancy in the NW quadrant between the flux ratios predicted by the double $\beta$-model and those inferred from the spectral normalisations may be due to the fact that the image decomposition assumes the presence of only two subclusters.

Several caveats should be kept in mind. As noted by \citet{2019ApJ...882...69G}, the optical subcluster decomposition is not unique and more complex merger geometries remain possible. In addition, the spectroscopic galaxy sampling is spatially non-uniform, particularly near the XRISM field of view and the cluster centre. Finally, mergers are expected to produce offsets between the ICM, galaxies, and dark matter following core passage, introducing further uncertainty into any direct component-to-component association. Nevertheless, the agreement between the XRISM line-of-sight velocities and the optical redshifts of Subclusters 1, 2, and 4 is remarkable.

\begin{figure*}[!ht]
    \centering

    \begin{subfigure}[t]{0.49\textwidth}
        \centering
        \includegraphics[width=\linewidth]{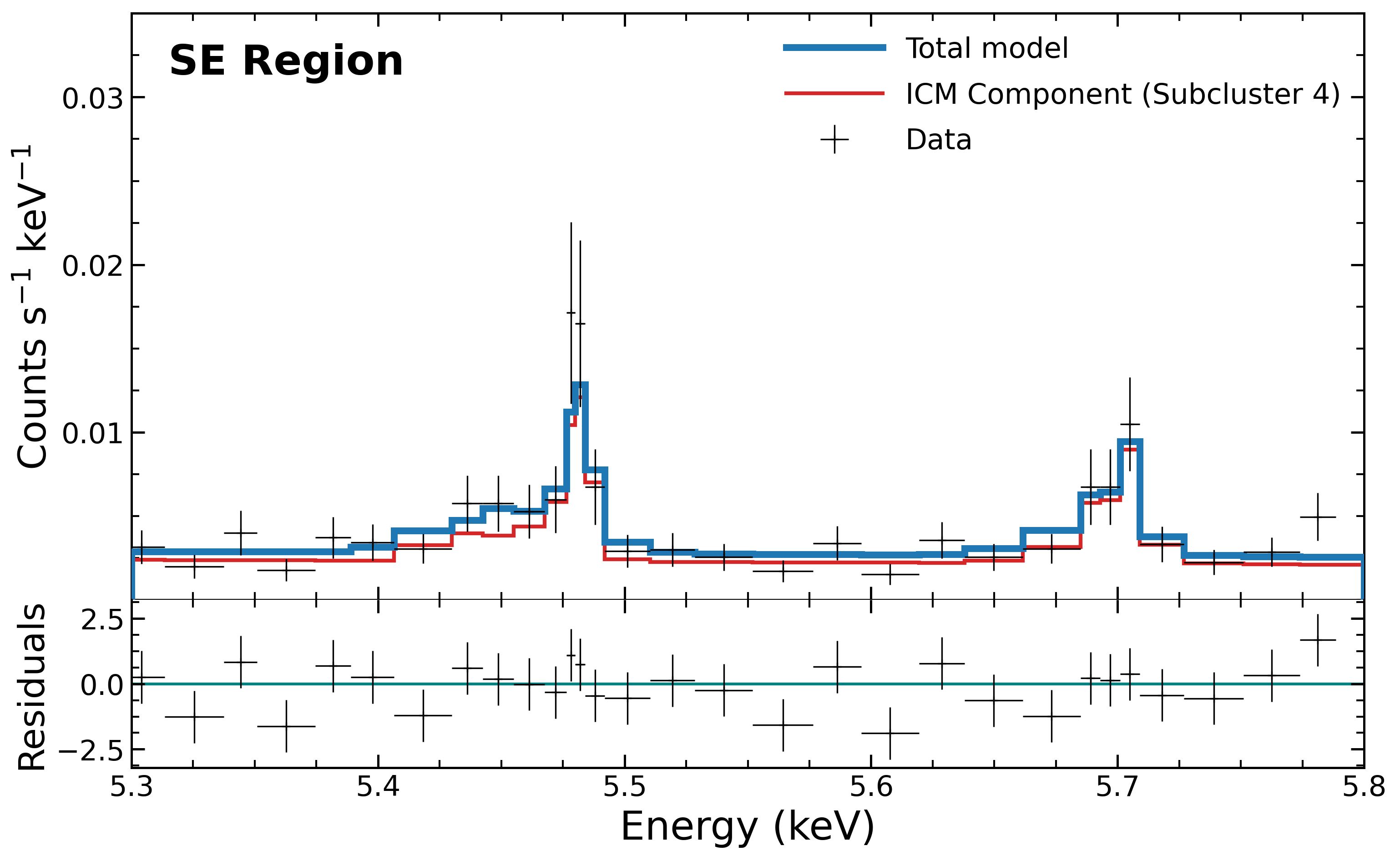}
        \label{SE_spectrum}
    \end{subfigure}
    \hfill
    \begin{subfigure}[t]{0.49\textwidth}
        \centering
        \includegraphics[width=\linewidth]{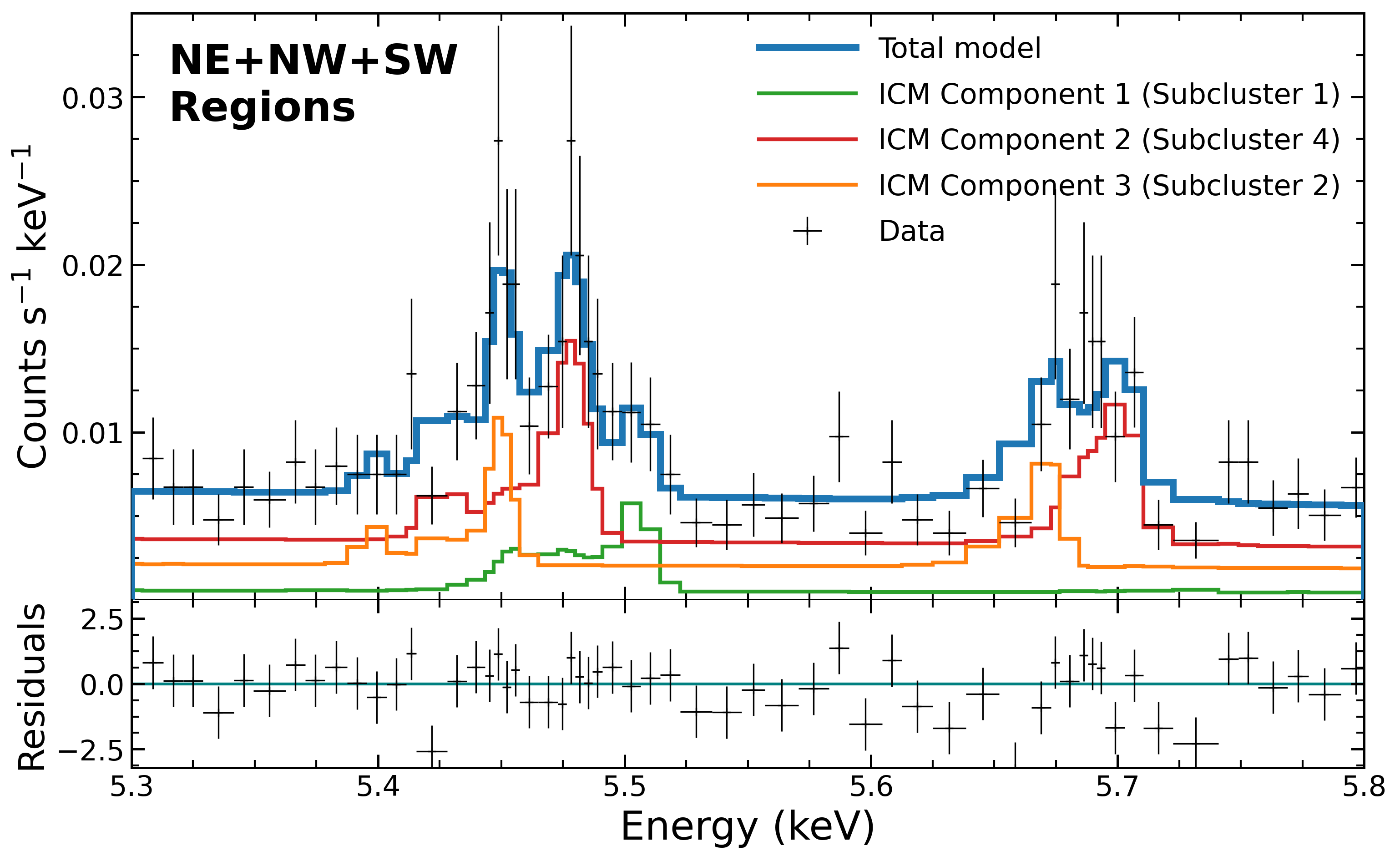}
        \label{north_spectrum}
    \end{subfigure}

    \caption{XRISM/Resolve spectra and best-fitting models plotted in the 5.3--5.8\,keV band for a two-region decomposition with the SE region (left) and the combined NE+NW+SW regions (right). The combined NE+NW+SW regions integrated spectra favours a \texttt{crossarf*tbabs*(bapec+bapec+bapec)} model. The three components that we detect have individual redshifts that are completely consistent with that of the three subclusters detected in previous weak-lensing and optical studies \citep{JeeEtAl2016, 2019ApJ...882...69G}. The corresponding optical subcluster associated with each ICM component detected by XRISM has been indicated alongside the component in the legend.}
    \label{resolve_spectral_components_beaut}
\end{figure*}

To better visualise this complex multi-velocity structure, we further group the XRISM spectra of the non-isothermal NE, NW, and SW quadrants, and compare their integrated spectrum with that of the SE region (see Fig.\,\ref{resolve_spectral_components_beaut}). Remarkably, we can indeed clearly observe XRISM/Resolve's ability to separate three line-of-sight ICM components in the Fe-K line complex in the combined three-quadrant spectrum, while the SE shows only one velocity component. We therefore model the SE region with a single thermal component and the combined three-quadrant spectrum with three components, accounting for spatial-spectral mixing. The best-fit parameters are reported in Table~\ref{SE_remaining_bestfit}. The three measured redshifts are consistent with those of optical Subclusters 1, 2, and 4. The dominant component is associated with Subcluster 4 ($z\sim0.223$), while the two additional components correspond to the northern subclusters ($z\sim0.217$ and $z\sim0.229$). The presence of these distinct velocity components confirms that XRISM is able to distinguish more than two merger constituents along our line of sight.

An early study by \cite{BruggenvanWeerenRottegering2012} had indeed concluded that a three-component merger was required to explain the unique morphology of the Toothbrush relic and the presence/positions of the two other radio relics within the cluster. Our XRISM results support a picture involving a merger between at least three components, however, we note that they differ from the collision geometry used in that simulation in two important ways. Firstly, the simulation-based study had predicted the third component to be located towards the east of the two main cores, which does not agree with our results. Instead, the additional X-ray component appears to lie towards the northern/nortwestern part of the system, consistent with the geometry later identified in \textit{Chandra} and weak-lensing studies \citep{vanWeerenEtAl2016Toothbrush, JeeEtAl2016}. Secondly, owing to the high polarisation of the main relic, the simulation by \cite{BruggenvanWeerenRottegering2012} also positioned the collision in the plane of the sky. 
As discussed before, we detect significant line-of-sight kinematic components, which suggests that the merger is more three-dimensional compared to that assumption.

\begin{table*}
\centering
\caption{Best-fit parameters for the two-region division of the XRISM FOV of the Toothbrush Cluster. We fit the SE region using a \texttt{crossarf$\times$tbabs$\times$bapec} ICM model, while the other spectrum is fit using a \texttt{crossarf$\times$tbabs$\times$(bapec+bapec+bapec)} model. This sub-array analysis allows us to observe XRISM's extraordinary capability to resolve/separate line-of-sight thermal and velocity components in the ICM.}
\label{SE_remaining_bestfit}
\renewcommand{\arraystretch}{1.6}
\resizebox{\textwidth}{!}{%
\begin{tabular}{lccccccc}
\hline
\hline
\textbf{\rule{0pt}{4ex}} 
& \shortstack{\rule{0pt}{4ex}\textbf{Temperature}\\\textbf{(keV)}} 
& \textbf{\rule{0pt}{4ex}Abundance} 
& \textbf{\rule{0pt}{4ex}Redshift} 
& \shortstack{\rule{0pt}{4ex}\textbf{Bulk Velocity}\\\textbf{($\mathrm{km\,s}^{-1}$)}} 
& \shortstack{\rule{0pt}{4ex}\textbf{$\sigma_v$}\\\textbf{($\mathrm{km\,s}^{-1}$)}} 
& \shortstack{\rule{0pt}{4ex}\textbf{Normalisation}\\\textbf{($\times10^{-3}$)}} 
& \textbf{\rule{0pt}{4ex}C-stat} \\
\hline

\textbf{SE Region}
& $9.21_{-0.36}^{+0.43}$
& $0.323_{-0.024}^{+0.026}$
& $0.2223_{-0.0003}^{+0.0003}$
& $-930_{-75}^{+75}$
& $210_{-70}^{+65}$
& $2.40_{-0.07}^{+0.08}$
& $1228$ \\

\hline

\textbf{NE+NW+SW Regions}
& & & & & & & \\

Component 1
& $3.08_{-0.33}^{+0.35}$
& $0.323$ (coupled)
& $0.2170_{-0.0007}^{+0.0009}$
& $-2225_{-170}^{+220}$
& $280_{-115}^{+210}$
& $2.10_{-0.36}^{+0.37}$
& $1243$ \\

Component 2
& $9.21$ (coupled to SE)
& $0.323$ (coupled)
& $0.2233_{-0.0007}^{+0.0003}$
& $-685_{-170}^{+75}$
& $325_{-75}^{+115}$
& $3.65_{-0.14}^{+0.16}$
& \\

Component 3
& $9.81_{-2.32}^{+1.15}$
& $0.323$ (coupled)
& $0.2295_{-0.0004}^{+0.0004}$
& $830_{-100}^{+100}$
& $185_{-75}^{+95}$
& $2.53_{-0.81}^{+0.35}$
& \\

\hline
\end{tabular}%
}
\end{table*}

This agreement between the ICM motions and galaxy redshifts also tells us about the merger's evolutionary stage to a certain extent. In a head-on merger where the pericenters have completely passed through one another, ram pressure can cause the collisional gas to kinematically decouple from the collisionless subhalo which creates substantial offsets between their velocities \citep{Zhang2019}. The fact that the ICM components show velocities broadly consistent with their associated optical subclusters could therefore indicate that the merger is not actually head-on or that some of the subhalos have not yet undergone their closest pericenter passage.

\subsection{Turbulent and Bulk Motion Support}

The Toothbrush Cluster presents one rare example where XRISM's high spectral resolution allows us to largely separate the intrinsic line broadening due to unresolved or random gas motion along the line-of-sight, and the bulk velocities that trace large-scale merger-driven flows. If we were to assume pure isotropic turbulence in the two line-of-sight components from the full field of view analysis, the three-dimensional turbulent Mach number,
\begin{equation}
    \mathcal{M}_{3D} = \frac{\sqrt{3}\sigma_v}{c_s},
\end{equation}
for the redshifted component ($c_s = 1332_{-177}^{+153}$\,$\mathrm{km\,s}^{-1}$) is $\mathcal{M}_{3D} = 0.32^{+0.14}_{-0.23}$. Similarly, for the blueshifted component ($c_s = 1542_{-59}^{+61}$\,$\mathrm{km\,s}^{-1}$), the Mach number is $\mathcal{M}_{3D} = 0.35^{+0.11}_{-0.07}$. Then the kinetic pressure fraction, given by \citep{Eckert2019},
\begin{equation}
    \frac{P_{\mathrm{kin}}}{P_{\mathrm{tot}}} = \frac{\mathcal{M}_{3D}^2}{\mathcal{M}_{3D}^2+ 3/\gamma},
\end{equation}
is $5.3^{+5.2}_{-4.9}\%$ for the redshifted and $6.3^{+4.3}_{-2.2}\%$ for the blueshifted component. Similarly, in the sub-array analysis, the main component has a 3D Mach number of $\mathcal{M}_{3D} = 0.23^{+0.03}_{ -0.08}$ and a kinetic pressure support fraction of only $P_{\rm kin}/P_{\rm tot} =2.9^{+0.9}_{-1.7}\%$. 

At first sight, such low levels of turbulent pressure support appear surprising for a system as dynamically disturbed as the Toothbrush Cluster. However, similar results have emerged from recent XRISM observations of other major mergers. The Coma Cluster shows non-thermal pressure fractions of $3.1\pm0.4\%$ and $3.3\pm0.8\%$ in its northern and southern pointings, respectively \citep{XRISMComa2025}, while Abell~754 exhibits values of $5.7^{+0.9}_{-0.8}\%$ and $2.4^{+0.8}_{-0.7}\%$ in its two observed regions \citep{OmiyaEtAl2025}. 

However, there is an important caveat: these estimates do not include the significant bulk velocity differences across (a) the two overlapping line-of-sight components or (b) between different regions of the field of view. While these merger-scale coherent bulk flows do not act as a local pressure component like isotropic turbulence, their associated kinetic energy density and ram pressure can strongly influence the dynamics \citep{OmiyaEtAl2025}. 

Another important observational consideration is that X-ray spectroscopy necessarily probes the emissivity-weighted line broadening, which may not be equivalent to the mass-weighted velocity dispersion of the ICM. The measured $\sigma_v$ is sensitive to the spatial scales that contribute most to the X-ray emission within a particular extraction region, while larger eddies are down-weighted because they affect fainter gas at larger projected distances along the line of sight. A common quantity used to assess the impact of this projection averaging is the so-called effective length $l_{eff}$, defined as the region along the line-of-sight from which 50\% of the X-ray emission originates \citep{ZhuravlevaEtAl2012}. Using the double-$\beta$ model in Section\,\ref{checking if vel of beta model are robust}, which can easily be converted into a 3D emissivity distribution, a crude estimate of $l_{eff}$ for the full FOV observation of the Toothbrush Cluster would be 205\,kpc, assuming a 30$^{\circ}$ angle of inclination between the two main subclusters. This means that turbulent velocities on scales larger than $\sim$200\,kpc do not significantly contribute to the observed line widths, possibly leading to an underestimation of the non-thermal pressure fraction. However, note that the effects of inhomogeneous and intermittent turbulence in the ICM can also potentially complicate the definition of a single meaningful characteristic scale (see, e.g., \cite{Vazza2026} for a discussion based on numerical simulations).

An additional ambiguity arises from the fact that XRISM detected multi-phase gas across three-quarters of the pointing. Since the best-fit model relied on tying the velocity broadening across regions given the existing statistics of the spectrum, it is not possible to determine how different thermal phases contribute to the measured broadening. This means that we can not interpret the measured $\sigma_v$ for the secondary component as evidence for strong turbulence within the cooler phase (NE and SW regions, which have a lower sound speed and hence higher Mach number), nor as weak turbulence within the hotter phase (NW region).

The contrast between the large bulk velocity offsets and comparatively small velocity dispersions provides additional clues about the distribution of kinetic power across spatial scales. Similar behaviour has been reported in other XRISM merger studies. In the Coma Cluster, the maximum line of sight velocity offset was observed to be about $730$\,$\mathrm{km\,s}^{-1}$ for the southern pointing, while the velocity dispersion was only $\sigma_v$\,\,$\sim$\,\,200\,$\mathrm{km\,s}^{-1}$ \citep{XRISMComa2025}. Likewise in Abell~1914, the two merging components exhibited a line-of-sight velocity separation of $\sim$1000\,$\mathrm{km\,s}^{-1}$ and a dispersion of only $\sim$200\,$\mathrm{km\,s}^{-1}$ \citep{Heinrich2025}. 

The Toothbrush Cluster displays an even more extreme version of this trend, with the bulk velocities reaching upwards of $\sim$2000\,$\mathrm{km\,s}^{-1}$ and velocity dispersions of only $\sim$200\,$\mathrm{km\,s}^{-1}$. In general, such measurements are consistent with a scenario where only $\sim$10--15 percent of the $PdV$ work done by the gas dynamics during mergers is channeled into turbulence \citep{2003ApJ...584..190F,Cassano2005}. The observed velocity structure could also indicate a velocity power spectrum steeper than Kolmogorov, as originally suggested for Coma \citep{XRISMComa2025}. However, using cosmological simulations, it has recently been argued that a spatially-intermittent and multi-scale distribution of turbulence with an underlying Kolmogorov-like spectrum can also reproduce the velocity measurements made by XRISM \citep{Vazza2026}. This complex distribution of motions can also affect the X-ray emissivity weighting and bias low the inferred total kinetic pressure fraction. 

Alternatively, the reason for the large bulk velocities compared to small line widths could be due to a young dynamical age of the merger. If indeed our updated age estimate of 0.2~Gyr presented in Sect.~\ref{section 5.1} is correct, this is substantially shorter than the expected large-eddy turnover time of $\sim1$\,Gyr (given by $t_\mathrm{eddy}\sim L/\sigma_v$), and would be consistent with much of the merger kinetic energy still residing in coherent large-scale flows rather than fully developed turbulence. A similar interpretation was recently discussed for the Coma Cluster as well, where the observed intracluster gas motions were attributed to a transient merger-driven phase \citep{Zhang2026Coma}.

This would in turn have important implications for the origin of the cluster's bright radio halo. If the turbulent cascade has not yet had time to propagate to smaller scales, turbulent reacceleration would be expected to be less efficient, and the remarkable uniformity of the radio spectral index would also be somewhat more difficult to explain. It is then possible that the halo could be sustained (at least in part) by turbulence generated during an earlier merger event. In addition, rather than having an intrinsically uniform spatial distribution, the observed radio spectral index may appear constant due to enhanced projection effects related to the merger inclination along the line of sight.


\section{Conclusions}\label{section 6}

We report a XRISM/Resolve study of the ICM velocity structure within the Toothbrush Cluster, using spatially resolved spectroscopy to measure the line-of-sight bulk velocities and velocity dispersions in the southern X-ray-bright region of the cluster. We have listed our main conclusions below. 

\begin{itemize}

\item The Toothbrush Cluster cannot be adequately described by a single ICM velocity component. In the full field-of-view analysis, the spectrum favours a two-component model with a line-of-sight velocity separation of $1635^{+165}_{-110}$ km\,s$^{-1}$. Despite this transonic offset, the individual components exhibit much smaller velocity dispersions of only $\sigma_v \sim 245$--310 km\,s$^{-1}$. 

\item We adopt this two-component description also in the spatially resolved quadrant analysis. The secondary component is significantly detected only in the NE, NW, and SW quadrants, whereas the primary component is present throughout the field of view. Together with an independent double $\beta$-model decomposition of the \textit{Chandra} image, this spatial distribution suggests that the primary component traces the southern subcluster, while the secondary component is associated with the northern subcluster. 

\item Large point-to-point variations are observed in the bulk velocities of both components, while the velocity dispersions decrease further to $\sigma_v \sim 210$--215 km\,s$^{-1}$. The secondary component is strongly blueshifted in the NE and SW regions ($v_\mathrm{bulk}\sim2155$--2520 km\,s$^{-1}$), but redshifted in the NW region ($v_\mathrm{bulk}\sim930$ km\,s$^{-1}$). The primary component remains blueshifted across the field, with bulk velocities ranging from $v_\mathrm{bulk}\sim$195--905 km\,s$^{-1}$. 

\item The secondary phase detected in the NE and SW regions has temperatures of 2--3.5 keV, substantially below the cluster's mean temperature ($kT_{\rm mean}\sim7.8\pm0.13$ keV; \citealt{Ogrean2013}), whereas the secondary component in the NW region has a temperature comparable to the cluster average. 

\item The distinct temperatures and redshifts measured for the northern component imply that it is not a single structure. By comparing our XRISM measurements with optical galaxy kinematics, we conclude that the northern system itself consists of two subhalos. XRISM is therefore able to separate three velocity components along the line of sight, corresponding to the southern subcluster and the two northern subhalos. The measured ICM redshifts are in good agreement with the subcluster decomposition obtained from the optical galaxy redshift distribution by \cite{2019ApJS..240...39G}. 

\item The observed velocity structure supports a post-core-passage merger geometry, consistent with previous weak-lensing and X-ray imaging studies \citep{JeeEtAl2016}. The large line-of-sight velocity offsets further imply that the merger has a substantial radial component, challenging interpretations based on radio polarization studies that favoured a geometry close to the plane of the sky. 

\item These measurements confirm that the Toothbrush Cluster is a highly disturbed and dynamically complex system. The combination of large bulk velocities, modest velocity dispersions, and strong spatial velocity gradients suggests that much of the merger kinetic energy remains on large spatial scales and that the system has not yet evolved into a fully developed turbulent state.

\end{itemize}

To summarise, our results indicate that the ICM velocity field in the Toothbrush Cluster is characterised by supersonic bulk flows and substantially smaller velocity dispersions within the multiple gas components detected by XRISM. This may suggest that only a relatively small fraction (of the order of $\sim$10\%) of the energy dissipated during collision is drained into turbulence on the scales probed by XRISM, which is approximately 200\,kpc. Alternatively, this may also indicate that sufficient time has not yet passed since the merger for a fully developed turbulence cascade to be set up, and that the interaction may be both younger and more three-dimensional than inferred based on previous radio and numerical simulation studies. Additional mapping of this target using high-resolution X-ray spectral imaging is essential to further understand the connection between ICM dynamics and particle acceleration in this complex system.


\section*{Data Availability}

The XRISM Toothbrush Cluster data set used in this work can be accessed by querying JAXA’s Data Archives and Transmission System
(DARTS).\footnote{\url{https://darts.isas.jaxa.jp/app/query/astroquery/basic.php?satelites=xrism}} The \textit{Chandra} dataset we use is available at \href{https://doi.org/10.25574/cdc.664}{https://doi.org/10.25574/cdc.664}.

\textit{Software:} \texttt{NUMPY} \citep{vanDerWalt2011}, \texttt{ASTROPY} \citep{Astropy2013, Astropy2018, Astropy2022}, \texttt{MATPLOTLIB} \citep{Hunter2007}, \texttt{SHERPA} \citep{2024ApJS..274...43S, 2009SciPy...51R, 2001SPIE.4477...76F}, \texttt{CIAO} \citep{Fruscione2006}, \texttt{HEASoft} \citep{1995ASPC...77..367B, 2014ascl.soft08004N}, \texttt{XSPEC} \citep{Arnaud1996}

\begin{acknowledgements}
      CZ acknowledges the support of the Czech Science Foundation (GACR) Junior Star grant number GM24-10599M.\\
\end{acknowledgements}

\bibliographystyle{bibtex/aa}
\bibliography{Toothbrush_AA_references}{}


\begin{appendix}

\clearpage
\FloatBarrier
\onecolumn

\section{Full FOV spectral fits using single- and multi-temperature models}\label{full fov full band spectra}

In Appendix\,\ref{full fov full band spectra}, we present a comparison between the full field of view XRISM/Resolve spectra (3--8\,keV) fitted with a single- (top) and two-component (bottom) thermal model respectively. In the first case, the \texttt{tbabs$\times$bapec} model provides a reasonable fit, however, it leaves some residuals around the dominant Fe-K lines and does not reproduce the line strengths and widths adequately. This suggests that a single thermal component is likely unable to encapsulate the complexities of the ICM. 

The \texttt{tbabs$\times$(bapec+bapec)} model gives a much better description of the data, especially the Fe-K line complex. The C-stat reduces by $\Delta$C-stat$=16$ for the second case, which is a meaningful improvement. The individual components in this case also have a much lower velocity dispersion ($\sim$245--310\,km/s) than the inflated line broadening we were observing in the case of a single-component model (750\,km/s). This suggests that the high dispersion was not due to unusually strong turbulence, but rather to account for the unresolved substructures along the line-of-sight.

\begin{figure}[!ht]
    \centering

    \begin{subfigure}{0.7\textwidth}
        \centering
        \includegraphics[width=\linewidth]{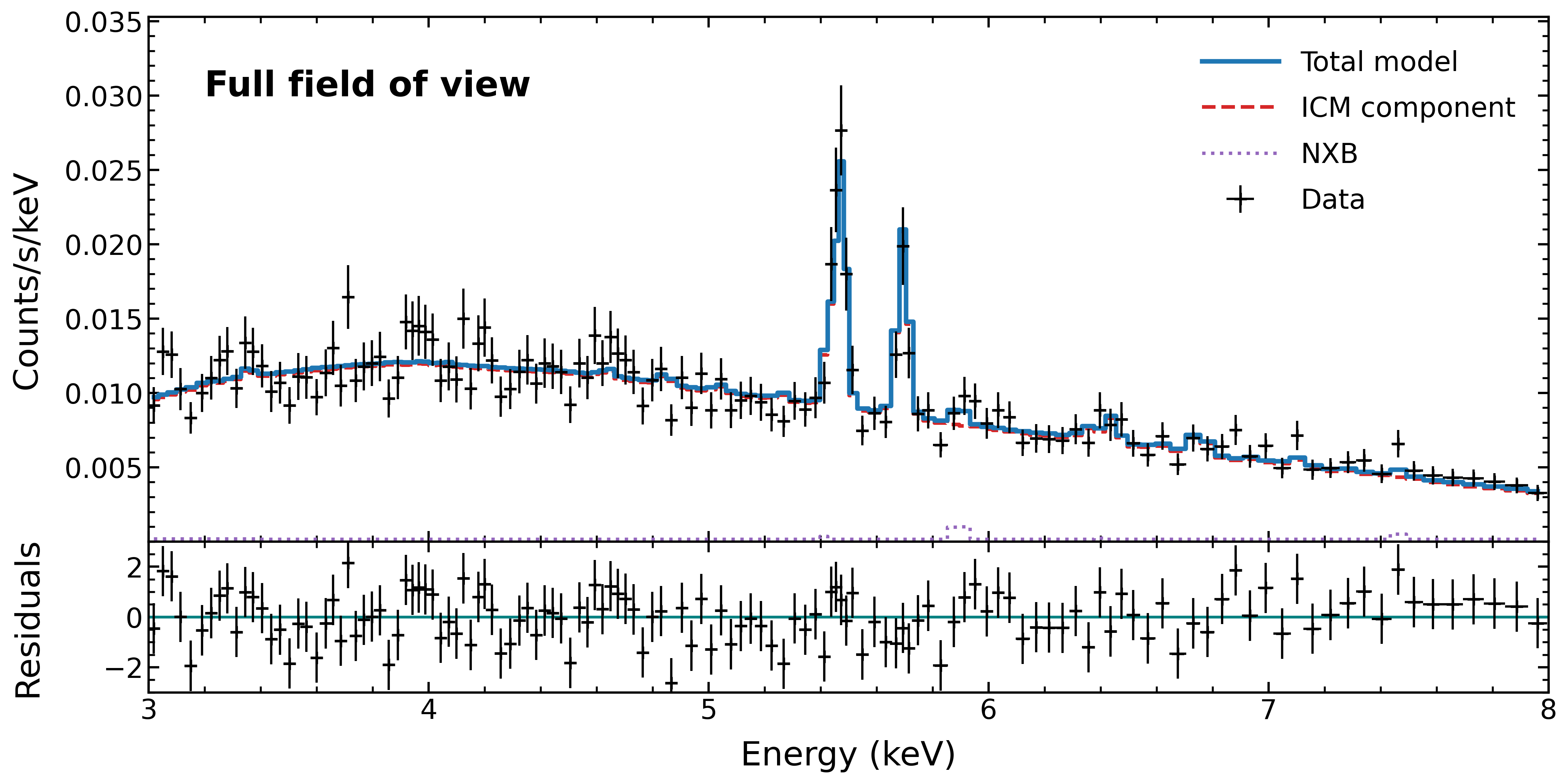}
        \label{fig:fullfov-1bapec}
    \end{subfigure}

    \vspace{0.5em}

    \begin{subfigure}{0.7\textwidth}
        \centering
        \includegraphics[width=\linewidth]{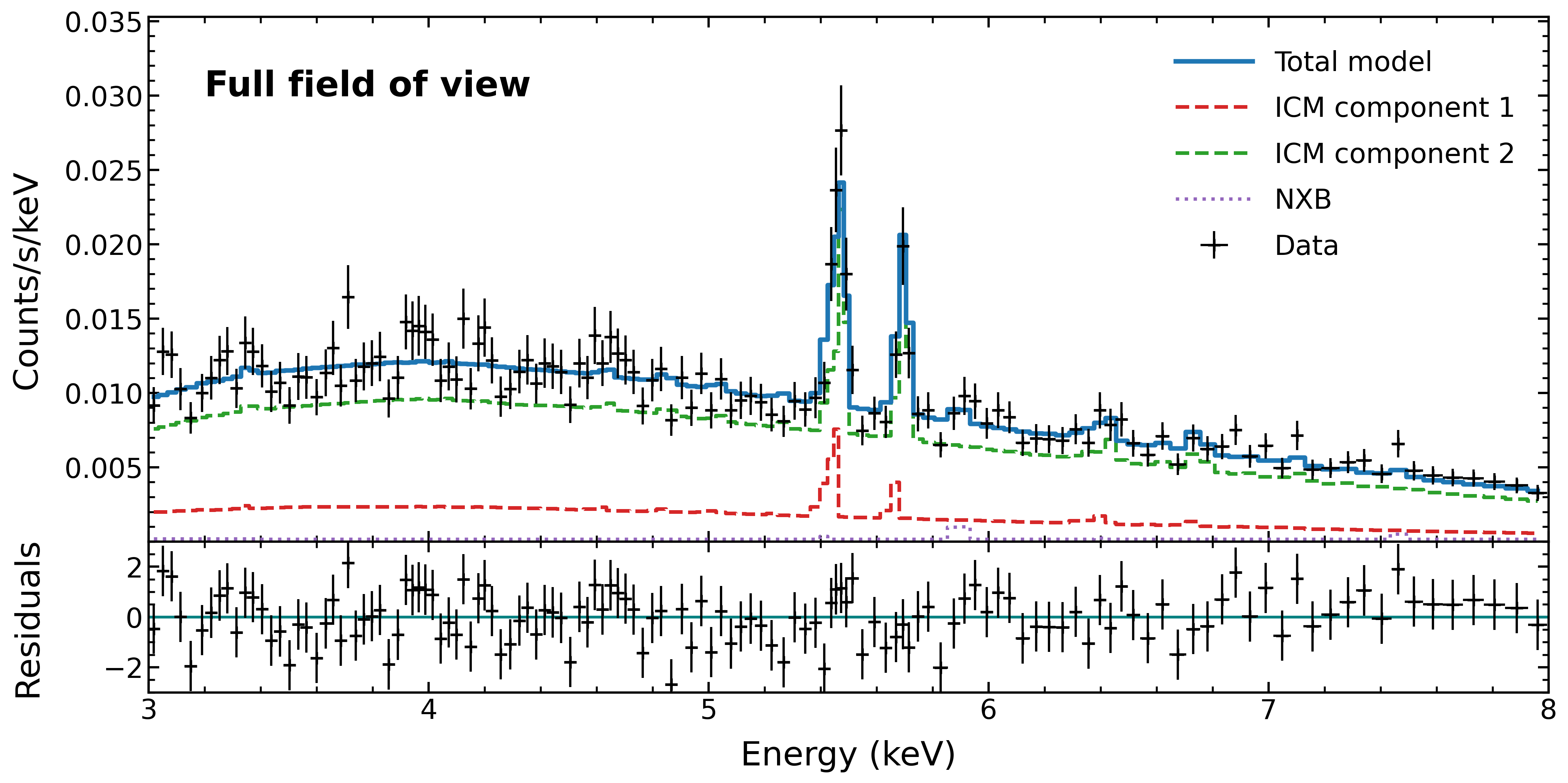}
        \label{fig:fullfov-2bapec}
    \end{subfigure}

    \caption{Comparison of single- and two-temperature models fit to the XRISM/Resolve full field of view spectrum of the Toothbrush cluster in the 3--8\,keV band. Top: The spectrum is fit using the \texttt{tbabs$\times$bapec} model. However, we observe that the widths and strengths of the dominant Fe-K lines are poorly fit. This model has a best-fit C-stat of 1354. Bottom: The spectrum is fit using the two-temperature \texttt{tbabs$\times$(bapec+bapec)} model. This model significantly improved the fitting of the dominant Fe-K lines. This model has a best-fit C-stat of 1338 ($\Delta$C-stat = 16).}
    \label{fig:fullfov-comparison}
\end{figure}

\section{The best-fit model for the full-array integrated spectrum when the temperatures of the two velocity components are tied}\label{fullarray temp tied}

While conducting the full-array analysis in Section\,\ref{full fov analysis}, we observe that the two detected ICM components have temperatures that are consistent within their 1$\sigma$ statistical uncertainties. This led us to tie the temperatures between the two and redo the two-component spectral fitting procedure. What we observe is that the velocity structure, i.e., the bulk velocities and dispersions, remain consistent across either case. This indicates that the need for two components along the line of sight arises, not due to a temperature contrast, but because of their substantial velocity separation. 

\begin{table}[!ht]
\centering
\caption{Best-fit parameters for the two-component ICM model (\texttt{tbabs$\times$(bapec+bapec)}), for the full field of view spectrum of the Toothbrush Cluster. The redshifts, bulk velocities, and velocity dispersions remain consistent when compared to the model presented in Table\,\ref{full field of best fit parameters for both models}. There is no significant change in the C-statistics either. We therefore conclude that the need for two ICM components arises from their kinematic separation rather than a temperature contrast.}
\label{full field of best fit parameters for two component model}
\renewcommand{\arraystretch}{1.6}
\resizebox{\textwidth}{!}{%
\begin{tabular}{lccccccc}
\hline
\hline
\textbf{\rule{0pt}{4ex}} 
& \shortstack{\rule{0pt}{4ex}\textbf{Temperature}\\\textbf{(keV)}} 
& \textbf{\rule{0pt}{4ex}Abundance} 
& \textbf{\rule{0pt}{4ex}Redshift} 
& \shortstack{\rule{0pt}{4ex}\textbf{Bulk Velocity}\\\textbf{($\mathrm{km\,s}^{-1}$)}} 
& \shortstack{\rule{0pt}{4ex}\textbf{$\sigma_v$}\\\textbf{($\mathrm{km\,s}^{-1}$)}} 
& \textbf{\rule{0pt}{4ex}Normalisation} 
& \shortstack{\rule{0pt}{4ex}\textbf{C-stat}\\\textbf{(Expected C-stat)}} \\
\hline
\textbf{2 Component Model} & & & & & & & \\
Component 1 
& $8.60_{-0.34}^{+0.39}$ 
& $0.28_{-0.03}^{+0.03}$ 
& $0.2294_{-0.0004}^{+0.0004}$ 
& $805_{-100}^{+100}$
& $190_{-105}^{+125}$ 
& $7.71_{-1.65}^{+1.89}\times10^{-3}$ 
& $1339~(1296 \pm 52)$ \\
Component 2 
& $8.60$ (coupled) 
& $0.28$ (coupled) 
& $0.2225_{-0.0002}^{+0.0003}$ 
& $-880_{-50}^{+75}$ 
& $305_{-55}^{+75}$ 
& $2.49_{-0.19}^{+0.16}\times10^{-2}$ 
& \\
\hline
\end{tabular}%
}
\end{table}


\section{Sub-array Analysis: Single-Temperature Fit}\label{single temp model quadrants}

For the sub-array analysis, we divide the field-of-view into quadrants as shown in Fig.\,\ref{Tooth_subregions}. The full field of view analysis already revealed the existence of atleast two components along the line-of-sight, and we hence adopt an equivalent model for the sub-array division as well. However, before we start fitting a two-component model, we fit each spectra with a single temperature model as an initial diagnostic. The best-fit model for this case has been presented in Table\,\ref{single_bapec_quadrant_bestfit}. The model is characterised by significant spatial variation in terms of its bulk velocities, as well as the high velocity dispersions measured in the NN and NW regions. Rather than attributing these high dispersions to unusually strong turbulence, we proceed to check if they arise from unresolved structures along our line-of-sight. 

\begin{table}[!ht]
\centering
\caption{Best-fit parameters for the quadrant sub-array analysis of the Toothbrush Cluster using a single-temperature \texttt{crossarf$\times$tbabs$\times$bapec} model in the full 3--8,keV band. The line-of-sight bulk velocities are computed from the redshifts using $v_{\rm bulk}=(z-z_\mathrm{tooth})\times c/(1+z_\mathrm{tooth})$, with $z_\mathrm{tooth}=0.226$.}
\label{single_bapec_quadrant_bestfit}

\renewcommand{\arraystretch}{1.6}

\resizebox{\textwidth}{!}{%
\begin{tabular}{lccccccc}
\hline
\hline
\textbf{\rule{0pt}{4ex}Region}
& \shortstack{\rule{0pt}{4ex}\textbf{Temperature}\\\textbf{(keV)}}
& \shortstack{\rule{0pt}{4ex}\textbf{Abundance}\\\textbf{($Z/Z_{\odot}$)}}
& \textbf{\rule{0pt}{4ex}Redshift}
& \shortstack{\rule{0pt}{4ex}\textbf{Bulk Velocity}\\\textbf{($\mathrm{km\,s}^{-1}$)}}
& \shortstack{\rule{0pt}{4ex}\textbf{$\sigma_v$}\\\textbf{($\mathrm{km\,s}^{-1}$)}}
& \shortstack{\rule{0pt}{4ex}\textbf{Normalisation}\\\textbf{($\times\,10^{-3}$)}}
& \shortstack{\rule{0pt}{4ex}\textbf{C-stat}\\\textbf{(Expected C-stat)}} \\
\hline

\textbf{SE}
& $8.83_{-0.49}^{+0.57}$
& $0.36_{-0.06}^{+0.06}$
& $0.2224_{-0.0004}^{+0.0003}$
& $-905_{-100}^{+75}$
& $225_{-60}^{+70}$
& $2.36_{-0.08}^{+0.08}$
& $4738~(4758\pm90)$ \\

\textbf{NE}
& $5.17_{-0.26}^{+0.27}$
& $0.18_{-0.05}^{+0.05}$
& $0.2171_{-0.0005}^{+0.0006}$
& $-2200_{-120}^{+145}$
& $360_{-105}^{+215}$
& $2.64_{-0.13}^{+0.13}$
& \\

\textbf{NW}
& $9.10_{-0.59}^{+0.60}$
& $0.40_{-0.07}^{+0.07}$
& $0.2275_{-0.0008}^{+0.0008}$
& $340_{-195}^{+195}$
& $640_{-120}^{+150}$
& $2.56_{-0.09}^{+0.09}$
& \\

\textbf{SW}
& $8.11_{-0.48}^{+0.53}$
& $0.24_{-0.05}^{+0.05}$
& $0.2226_{-0.0003}^{+0.0004}$
& $-860_{-75}^{+100}$
& $160_{-90}^{+100}$
& $2.36_{-0.09}^{+0.09}$
& \\

\hline
\end{tabular}%
}
\end{table}

\section{Comparison of the best-fit models before and after tying temperatures of the southern component across the four quadrants subregions}\label{before tying temepratures of 3+4 model}

\begin{figure*} 
    \centering
    \makebox[0.8\textwidth][c]{%
        \includegraphics[width=0.9\textwidth]{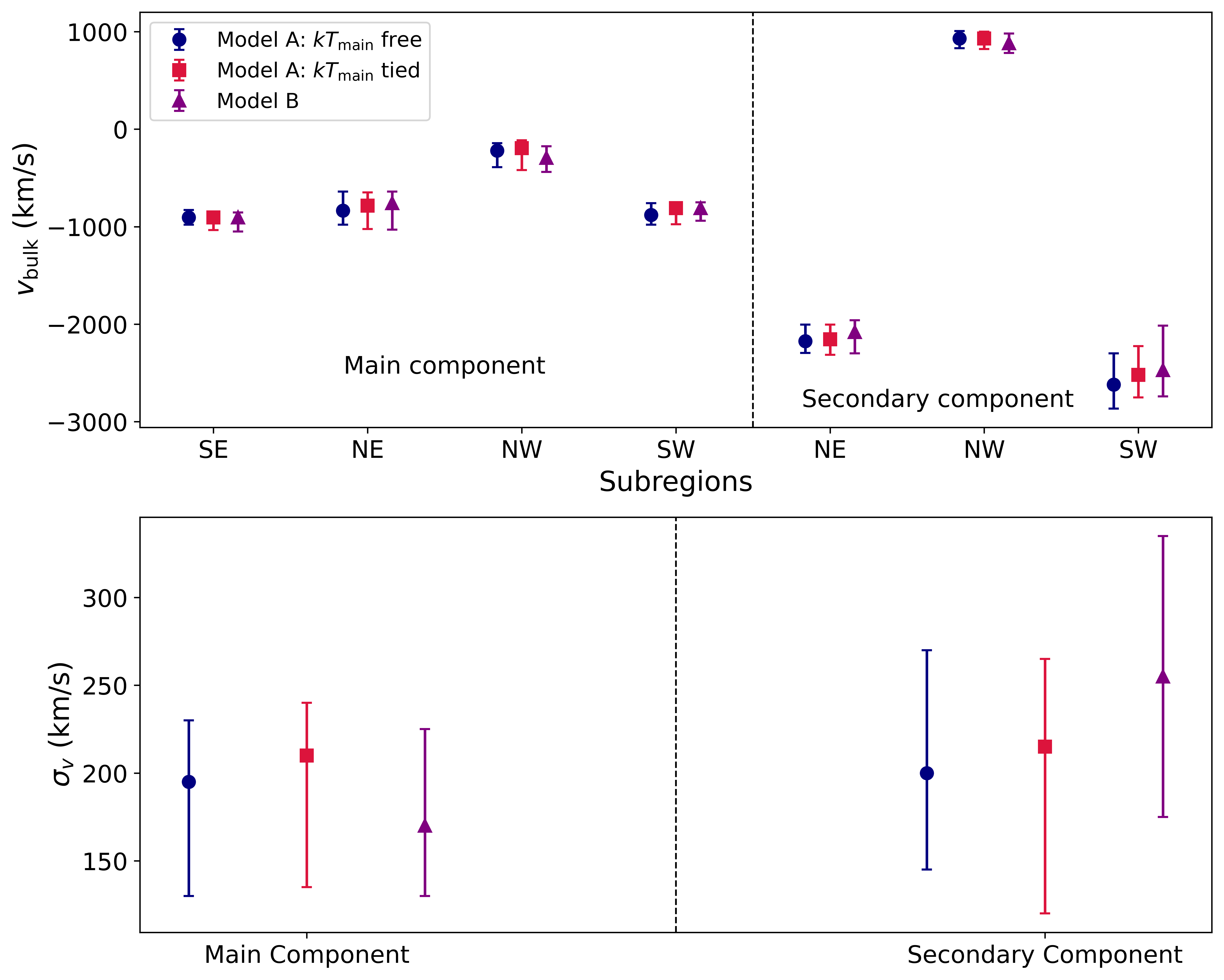}}
    \caption{The comparison of bulk velocities (top) and velocity dispersions (bottom) obtained in model\,A before and after tying the temperatures of the main component across all subregions. The blue markers show the best-fit values of velocities when the main component's temperature parameters, $kT_{\mathrm{main}}$, are free, whereas the red markers show the fit values when the temperatures are tied. We therefore conclude that tying the temperatures of the main component across the quadrants does not affect our velocity measurements. We also plot the velocity measurements obtained from model\,B (purple markers). We exclude the model B velocity measurement in the SE region since its value was not constrained.}
    \label{before and after temp are tied vel map}
\end{figure*}

As we carry out the spectral fitting of the sub-array spectra, we initially allow the temperatures to vary freely, along with the redshifts and model normalisations. However, it is soon found that the temperatures of the main component across all four subregions are mutually consistent. The best-fit model at this stage is reported in Table\,\ref{before tying temp of southern comop}. In Fig.\,\ref{before and after temp are tied vel map}, we plot and compare the velocity measurements across the two cases. $kT_{\mathrm{main}}$ denotes the temperatures of the brighter main component. The blue markers in the figure show velocity measurements when the temperatures are allowed to vary freely, while the red markers show results when the temperatures are tied across all subregions. We also include the velocity measurements from model\,B (represented by the purple-coloured markers), which has been discussed in Appendix\,\ref{model B}.

\begin{table}[!ht]
\caption{Best-fit parameters for the quadrant configuration of the Toothbrush Cluster using the two-temperature \texttt{crossarf$\times$tbabs(bapec+bapec)} model in the full 3--8\,keV band before the main component's temperatures are tied. The line-of-sight bulk velocity was computed from the redshift using the formula $v_{\rm bulk}=(z-z_\mathrm{tooth})\times c/(1+z_\mathrm{tooth})$.}
\label{before tying temp of southern comop}

\centering
\renewcommand{\arraystretch}{1.4}

\begin{tabular}{lcc}
\hline\hline
Parameter (Unit)
& \shortstack{Main\\component}
& \shortstack{Secondary\\component} \\
\hline

Abundance ($Z/Z_\odot$)
& $0.34^{+0.03}_{-0.03}$
& $0.41^{+0.07}_{-0.05}$ \\

$\sigma_v$ ($\mathrm{km\,s}^{-1}$)
& $195^{+35}_{-65}$
& $200^{+70}_{-55}$ \\
\hline

\multicolumn{3}{l}{\textbf{SE Region}} \\

$kT$ (keV)
& $9.14^{+0.44}_{-0.62}$
& -- \\

Redshift
& $0.2224^{+0.0003}_{-0.0003}$
& -- \\

$v_{\rm bulk}$ ($\mathrm{km\,s}^{-1}$)
& $-905^{+75}_{-75}$
& -- \\

Norm ($10^{-3}$)
& $2.33^{+0.08}_{-0.07}$
& -- \\
\hline

\multicolumn{3}{l}{\textbf{NE Region}} \\

$kT$ (keV)
& $10.90^{+3.03}_{-1.43}$
& $3.40^{+0.26}_{-0.29}$ \\

Redshift
& $0.2227^{+0.0008}_{-0.0006}$
& $0.2172^{+0.0007}_{-0.0005}$ \\

$v_{\rm bulk}$ ($\mathrm{km\,s}^{-1}$)
& $-835^{+195}_{-145}$
& $-2175^{+170}_{-120}$ \\

Norm ($10^{-3}$)
& $1.02^{+0.10}_{-0.06}$
& $1.49^{+0.18}_{-0.16}$ \\
\hline

\multicolumn{3}{l}{\textbf{NW Region}} \\

$kT$ (keV)
& $11.78^{+2.34}_{-1.66}$
& $6.69^{+0.77}_{-0.66}$\\

Redshift
& $0.2252^{+0.0003}_{-0.0007}$
& $0.2299^{+0.0003}_{-0.0004}$ \\

$v_{\rm bulk}$ ($\mathrm{km\,s}^{-1}$)
& $-220^{+75}_{-170}$
& $930^{+75}_{-100}$ \\

Norm ($10^{-3}$)
& $1.41^{+0.08}_{-0.08}$
& $1.21^{+0.11}_{-0.09}$\\
\hline

\multicolumn{3}{l}{\textbf{SW Region}} \\

$kT$ (keV)
& $11.87^{+1.87}_{-1.24}$
& $1.70^{+0.14}_{-0.13}$ \\

Redshift
& $0.2225^{+0.0005}_{-0.0004}$
& $0.2154^{+0.0013}_{-0.0010}$ \\

$v_{\rm bulk}$ ($\mathrm{km\,s}^{-1}$)
& $-880^{+120}_{-100}$
& $-2620^{+320}_{-245}$ \\

Norm ($10^{-3}$)
& $1.72^{+0.07}_{-0.07}$
& $2.37^{+0.52}_{-0.46}$ \\
\hline

C-stat (Expected C-stat)
& \multicolumn{2}{c}{$4714\, (4762\pm 90)$} \\

DOF
& \multicolumn{2}{c}{4166} \\
\hline

\end{tabular}
\end{table}

When we allow the temperatures of the southern component across the four subregions to vary freely, the resulting bulk and turbulent velocity measurements were consistent with those obtained when the temperatures are tied (see Fig.\,\ref{before and after temp are tied vel map}). The largest change was observed in the SW region, where there was a difference of 50\,$\mathrm{km\,s}^{-1}$ between the bulk velocities of the main component; the shifts in other regions were 15--30\,$\mathrm{km\,s}^{-1}$. The bulk velocity uncertainties also remain broadly comparable between the two cases. Similarly, the velocity dispersion of the main component changes slightly from $\sigma_v=195$\,$\mathrm{km\,s}^{-1}$ to 210\,$\mathrm{km\,s}^{-1}$ when its temperatures are tied. That of the secondary component rises as well, from $200^{+70}_{-55}$\,$\mathrm{km\,s}^{-1}$ to $215^{+50}_{-95}$\,$\mathrm{km\,s}^{-1}$. We therefore conclude that tying the temperatures of the main component does not alter the velocity structure in any substantial manner; it simply reduces the number of free parameters in our model.

\section{Quadrant sub-array full band spectra for Model A}

\begin{figure*}[!ht]
    \centering
    \includegraphics[width=0.495\textwidth]{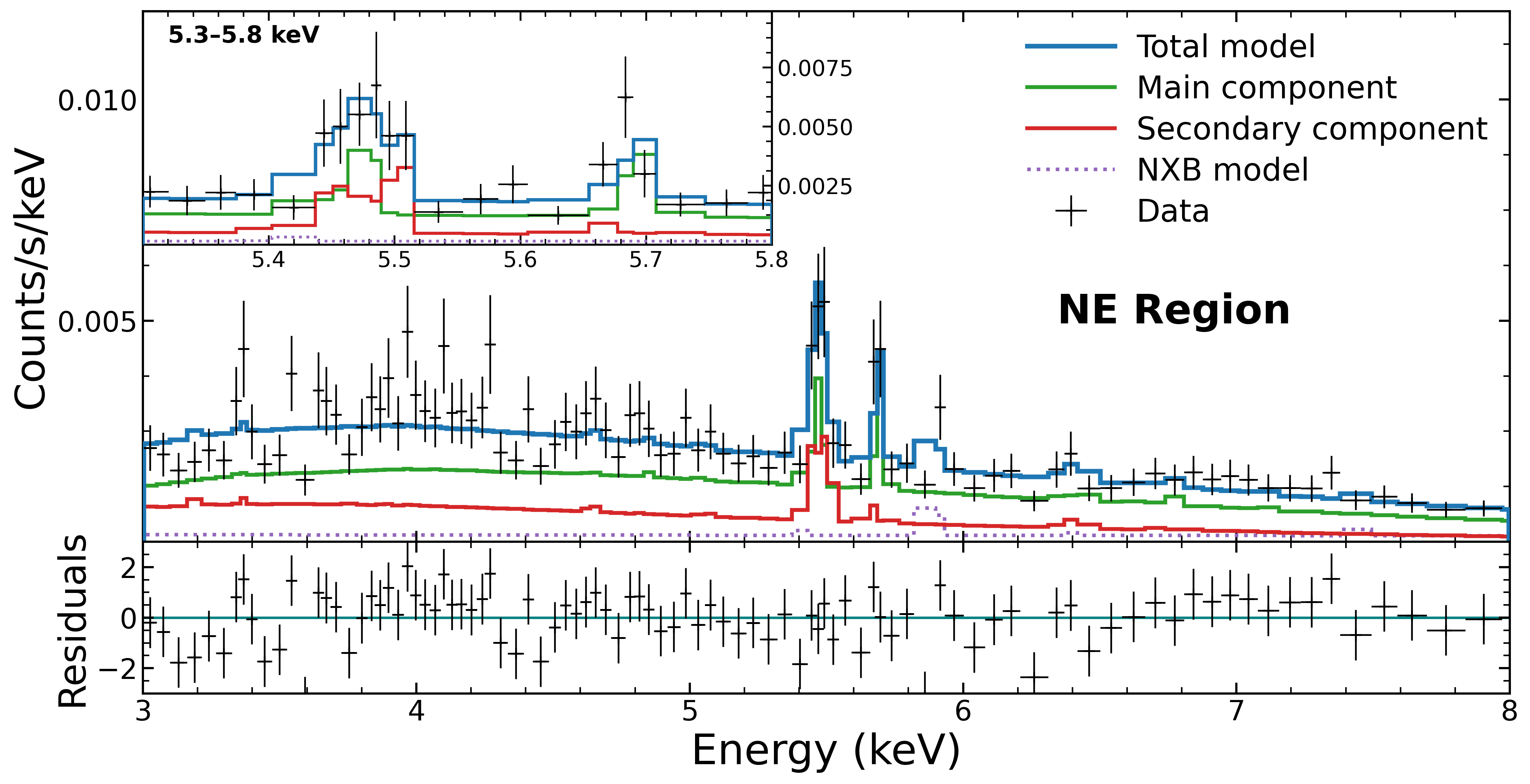}
    \hfill
    \includegraphics[width=0.495\textwidth]{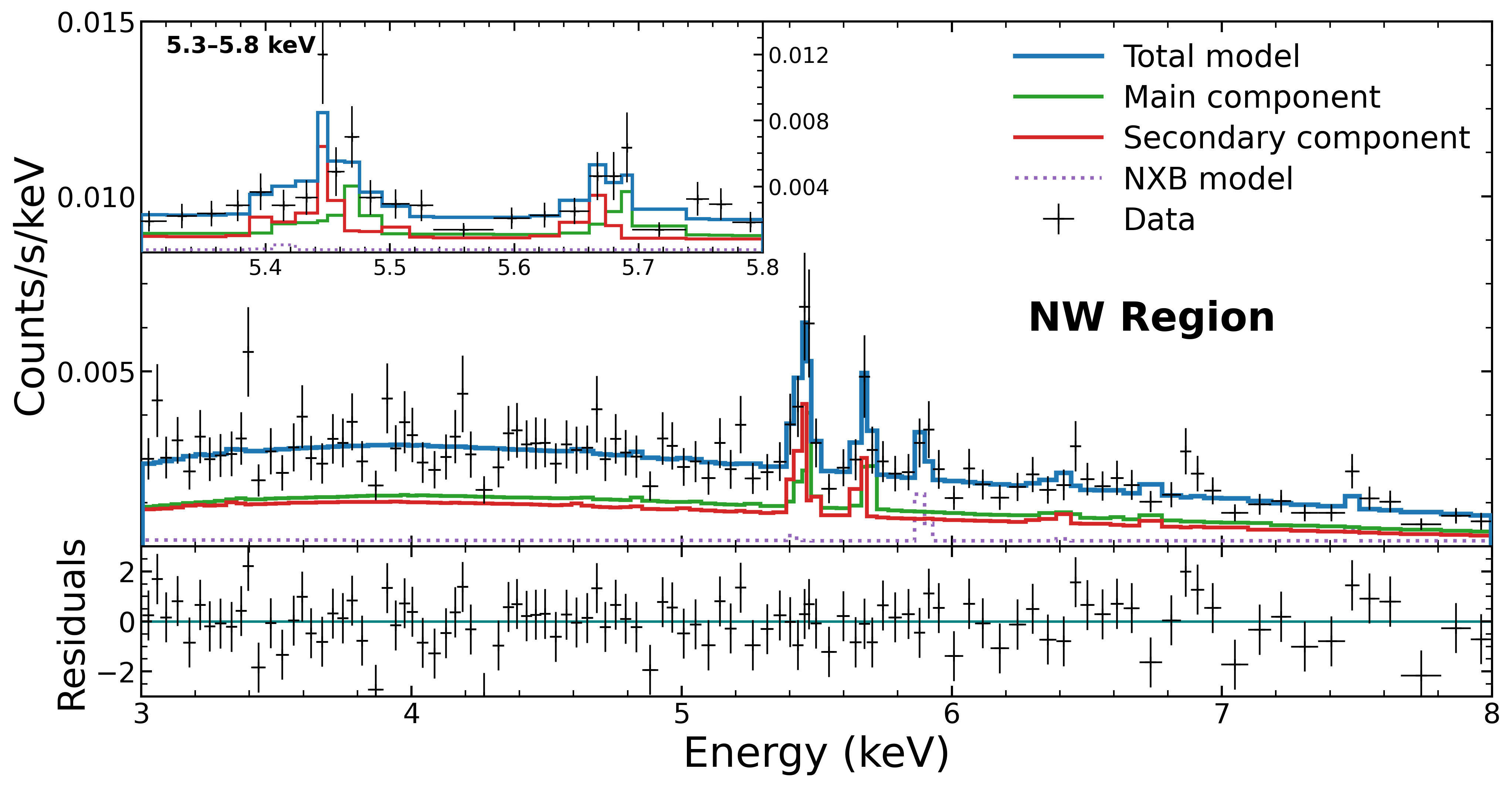}

    \vspace{0.02\textheight}

    \includegraphics[width=0.495\textwidth]{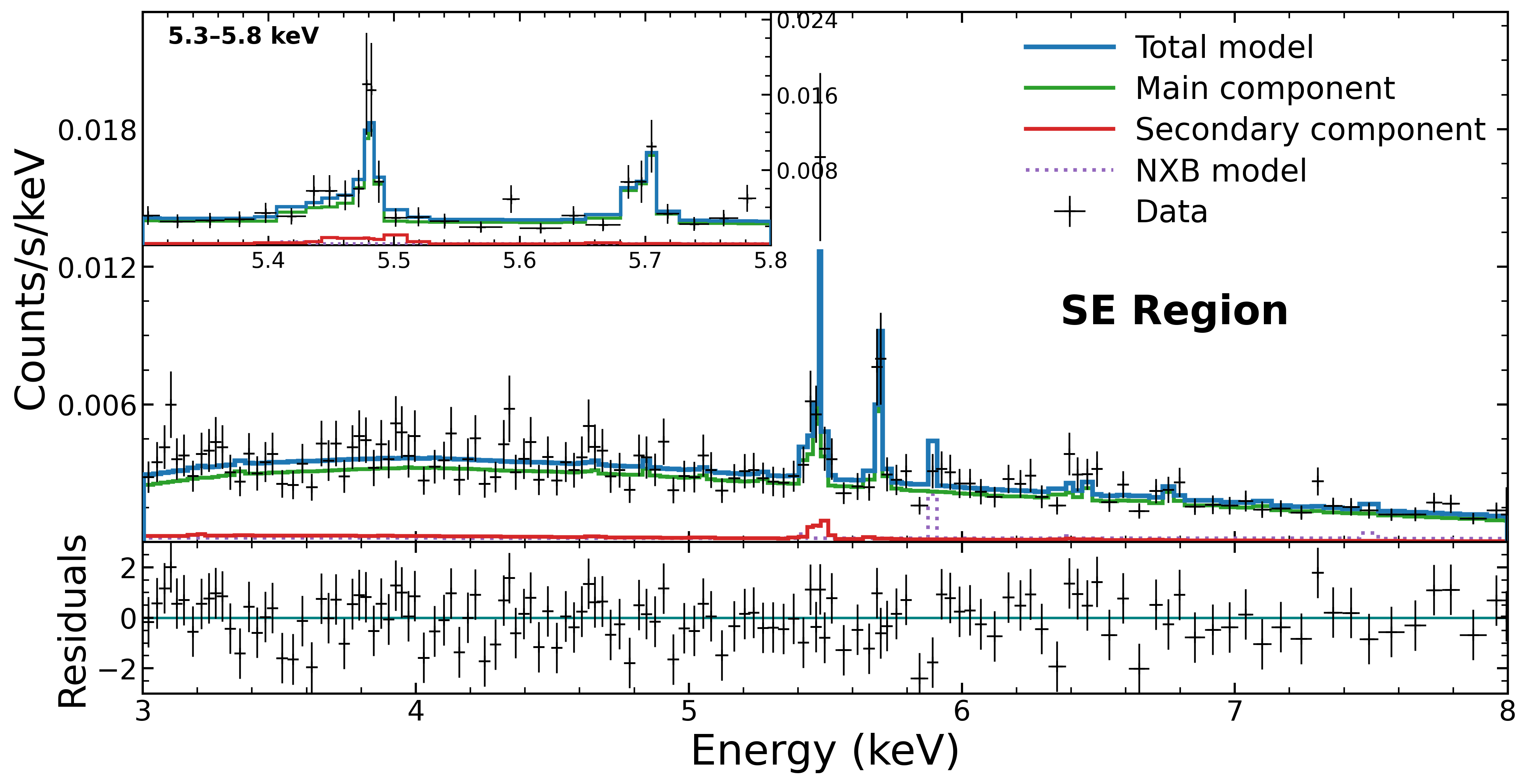}
    \hfill
    \includegraphics[width=0.495\textwidth]{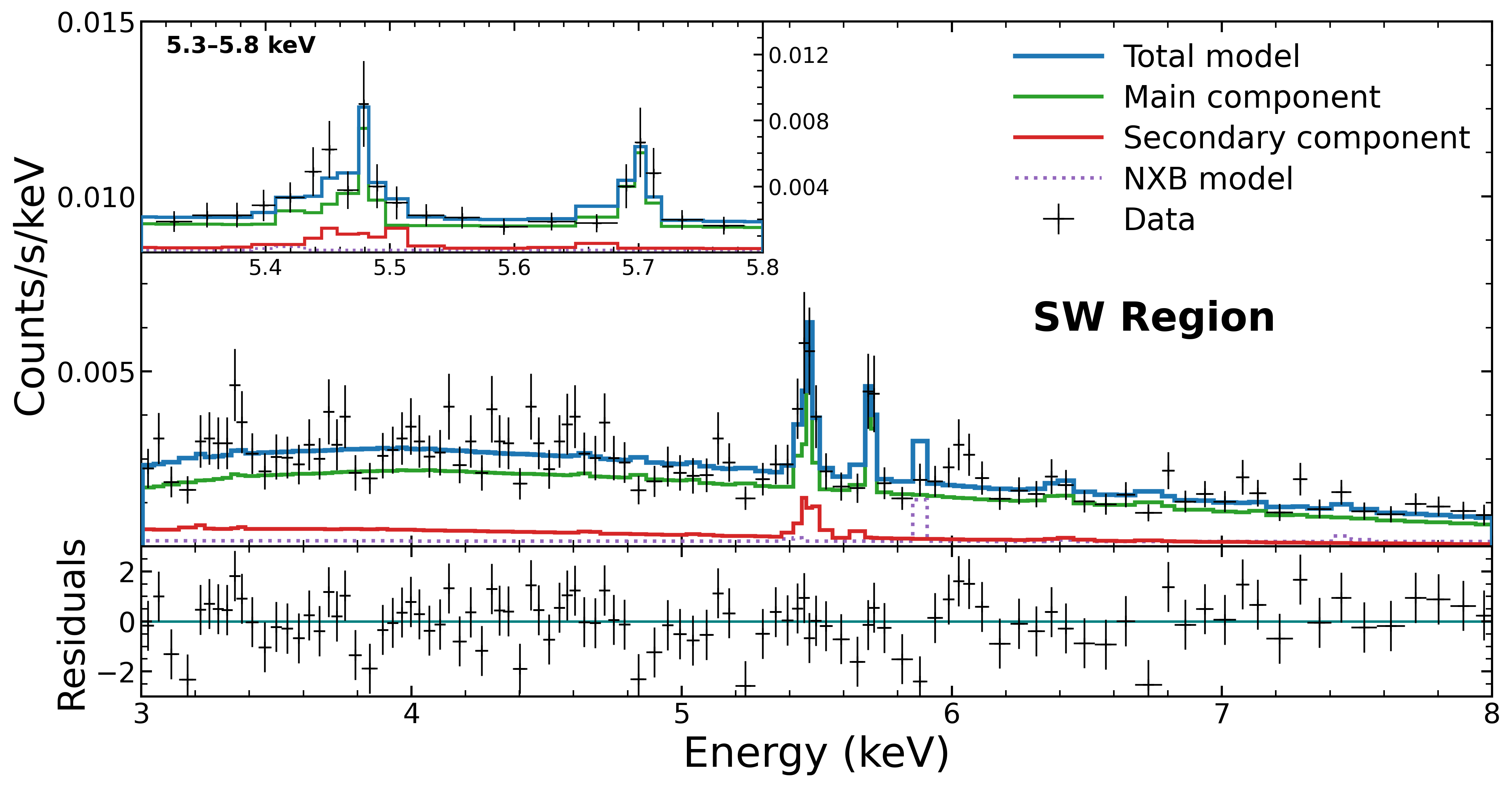}

    \caption{Best-fit two-temperature \texttt{crossarf$\times$tbabs(bapec+bapec)} spectral models for the four XRISM/Resolve sub-regions of the Toothbrush cluster over the 3--8\,keV band. In the SE region, we observe there to be only one statistically significant ICM component, whereas we detect two components in all other regions. The most significantly detected lines correspond to Fe\,XXV\,He$\alpha$ and Fe\,XXVI\,Ly$\alpha$ respectively. This model has a C-stat of 4715.}
    \label{fig:four_panel}
\end{figure*}

In this Appendix, we present the full-band spectra and the best-fit two-component model for each of the sub-array quadrants discussed in Section\,\ref{spatial dist of vel}. On the top left corner of each plot in Fig.\,\ref{fig:four_panel}, we also provide a zoomed-in view of each spectrum in the 5.3--5.8\,keV energy range, which allows us to observe the Fe-K line profiles more clearly.

\section{Impact of PSF calibration uncertainties on our best-fit model}\label{appendix sys unc data}

We test the robustness of our results from model\,A against the calibration uncertainties in the XRISM XMA and its point spread function. We do so by rescaling the cross-ARFs for each quadrant and redoing the spectral fitting procedure. The best-fit model obtained after rescaling the cross-ARFs by $\pm$30\% is presented in Table\,\ref{tab:quadrant_best_fit_crossarf_systematics_modelA}. The  velocity measurements in both cases remain consistent with that of model A within their 1$\sigma$ uncertainties. 

\begin{table*}[p]
\caption{Best-fit parameters for the quadrant configuration of the Toothbrush Cluster using the two-temperature \texttt{crossarf$\times$tbabs(bapec+bapec)} model. 
All the cross-region ARF files for model\,A have been scaled by $\pm30\%$. All parameters in both models are consistent with the fiducial model within their $1\sigma$ uncertainties.}
\label{tab:quadrant_best_fit_crossarf_systematics_modelA}

\centering
\renewcommand{\arraystretch}{1.55}
\setlength{\tabcolsep}{10pt}

\begin{tabular}{l|cc|cc}
\hline\hline
\noalign{\vskip 5pt}

&
\multicolumn{2}{c|}{\textbf{Cross-ARF: $-30\%$}}
&
\multicolumn{2}{c}{\textbf{Cross-ARF: $+30\%$}}
\\[5pt]

Parameter (Unit)
& \shortstack{Main\\component}
& \shortstack{Secondary\\component}
& \shortstack{Main\\component}
& \shortstack{Secondary\\component}
\\[3pt]

\hline
\noalign{\vskip 3pt}

$kT$ (keV)
& $9.48^{+0.27}_{-0.37}$
& --
& $9.50^{+0.37}_{-0.32}$
& --
\\

Abundance ($Z/Z_\odot$)
& $0.31^{+0.03}_{-0.03}$
& $0.38^{+0.04}_{-0.08}$
& $0.31^{+0.03}_{-0.03}$
& $0.40^{+0.09}_{-0.05}$
\\

$\sigma_v$ ($\mathrm{km\,s}^{-1}$)
& $195^{+35}_{-65}$
& $210^{+65}_{-75}$
& $190^{+35}_{-65}$
& $195^{+55}_{-90}$
\\[2pt]

\hline
\multicolumn{1}{l|}{\textbf{SE region}}
& \multicolumn{2}{l|}{}
& \multicolumn{2}{l}{}
\\[2pt]

$kT$ (keV)
& $9.48$ (coupled)
& --
& $9.50$ (coupled)
& --
\\

Redshift
& $0.2224^{+0.0002}_{-0.0004}$
& --
& $0.2223^{+0.0003}_{-0.0005}$
& --
\\

$v_{\rm bulk}$ ($\mathrm{km\,s}^{-1}$)
& $-905^{+50}_{-100}$
& --
& $-930^{+75}_{-120}$
& --
\\

Norm ($10^{-3}$)
& $2.30^{+0.05}_{-0.09}$
& --
& $2.33^{+0.13}_{-0.06}$
& --
\\[2pt]

\hline
\multicolumn{1}{l|}{\textbf{NE region}}
& \multicolumn{2}{l|}{}
& \multicolumn{2}{l}{}
\\[2pt]

$kT$ (keV)
& $9.48$ (coupled)
& $3.30^{+0.24}_{-0.27}$
& $9.50$ (coupled)
& $3.06^{+0.25}_{-0.41}$
\\

Redshift
& $0.2228^{+0.0004}_{-0.0008}$
& $0.2174^{+0.0005}_{-0.0007}$
& $0.2227^{+0.0008}_{-0.0007}$
& $0.2175^{+0.0005}_{-0.0008}$
\\

$v_{\rm bulk}$ ($\mathrm{km\,s}^{-1}$)
& $-810^{+100}_{-195}$
& $-2130^{+120}_{-170}$
& $-835^{+195}_{-170}$
& $-2105^{+120}_{-195}$
\\

Norm ($10^{-3}$)
& $1.20^{+0.10}_{-0.06}$
& $1.53^{+0.12}_{-0.21}$
& $1.13^{+0.06}_{-0.22}$
& $1.46^{+0.19}_{-0.31}$
\\[2pt]

\hline
\multicolumn{1}{l|}{\textbf{NW region}}
& \multicolumn{2}{l|}{}
& \multicolumn{2}{l}{}
\\[2pt]

$kT$ (keV)
& $9.48$ (coupled)
& $8.01^{+0.70}_{-0.81}$
& $9.50$ (coupled)
& $8.13^{+1.06}_{-0.74}$
\\

Redshift
& $0.2247^{+0.0006}_{-0.0003}$
& $0.2299^{+0.0004}_{-0.0004}$
& $0.2252^{+0.0004}_{-0.0008}$
& $0.2298^{+0.0003}_{-0.0004}$
\\

$v_{\rm bulk}$ ($\mathrm{km\,s}^{-1}$)
& $-345^{+145}_{-75}$
& $930^{+100}_{-100}$
& $-220^{+100}_{-195}$
& $905^{+75}_{-100}$
\\

Norm ($10^{-3}$)
& $1.14^{+0.08}_{-0.08}$
& $1.49^{+0.07}_{-0.10}$
& $1.16^{+0.16}_{-0.07}$
& $1.52^{+0.15}_{-0.08}$
\\[2pt]

\hline
\multicolumn{1}{l|}{\textbf{SW region}}
& \multicolumn{2}{l|}{}
& \multicolumn{2}{l}{}
\\[2pt]

$kT$ (keV)
& $9.48$ (coupled)
& $2.35^{+0.37}_{-0.30}$
& $9.50$ (coupled)
& $2.32^{+0.38}_{-0.40}$
\\

Redshift
& $0.2226^{+0.0003}_{-0.0004}$
& $0.2159^{+0.0011}_{-0.0009}$
& $0.2227^{+0.0003}_{-0.0005}$
& $0.2157^{+0.0012}_{-0.0008}$
\\

$v_{\rm bulk}$ ($\mathrm{km\,s}^{-1}$)
& $-855^{+75}_{-100}$
& $-2495^{+270}_{-220}$
& $-835^{+75}_{-120}$
& $-2545^{+295}_{-195}$
\\

Norm ($10^{-3}$)
& $1.93^{+0.08}_{-0.06}$
& $0.96^{+0.27}_{-0.23}$
& $1.92^{+0.07}_{-0.11}$
& $0.95^{+0.31}_{-0.29}$
\\[3pt]

\hline
\noalign{\vskip 2pt}

C-stat
& \multicolumn{2}{c|}{$4716$}
& \multicolumn{2}{c}{$4717$}
\\

Expected C-stat
& \multicolumn{2}{c|}{$4762\pm91$}
& \multicolumn{2}{c}{$4762\pm91$}
\\

DOF
& \multicolumn{2}{c|}{4168}
& \multicolumn{2}{c}{4168}
\\[2pt]

\hline
\end{tabular}
\end{table*}

\begin{figure*}
    \centering
    \makebox[0.9\textwidth][c]{%
        \includegraphics[width=0.7\textwidth]{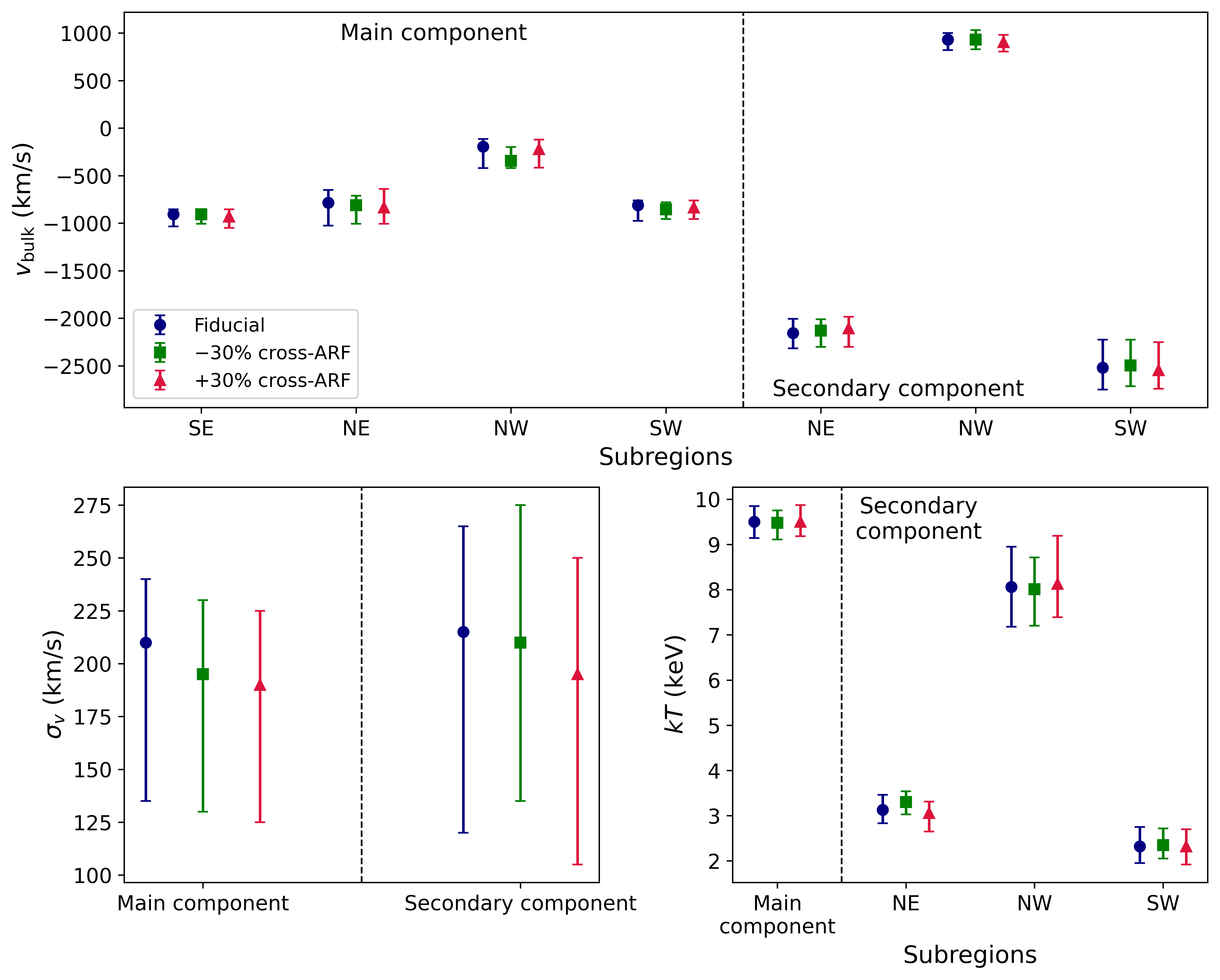}}
    \caption{Comparison of best-fit bulk velocity, velocity dispersion, and temperature measurements obtained for the fiducial `model\,A' and the models obtained when the cross-ARFs are shifted/scaled by $\pm 30\%$. The normalisations in model\,A are allowed to vary freely. We observe that all parameters are consistent with each other, within their $1\sigma$ uncertainties, across the three cases.}
    \label{contours_xray_on_chandra_rough}
\end{figure*}

\section{Best-fit model with the component normalisations fixed to the flux ratios predicted by the double $\beta$-model (`Model B')}\label{model B}

Based on the image analysis of Section\,\ref{checking if vel of beta model are robust}, we repeat the sub-array spectral analysis by fixing the normalisations of the two components to the flux ratios predicted by the double $\beta$-model, integrated over the respective sky regions (see Table \ref{double beta contributions}). The best-fit model obtained using this procedure, model\,B, is reported in Table \ref{fixed norm from beta model spectral fit}. 

As evident from Table\,\ref{fixed norm from beta model spectral fit}, the velocity structure of the two subclusters remains similar to what was obtained when we allow the normalisations of the two components to vary freely, as in model\,A (for the latter, see Table\,\ref{quadrant_best_fit_new_with_region_kt}). Across all four sub-regions, the bulk velocities remain essentially unchanged between the two cases. In the SE region, the spectrum only required a single component in model\,A, while in model B the northern component retains a low but non-zero normalisation, leading its bulk velocity to be poorly constrained.

\begin{table}[!ht]
\caption{Best-fit parameters for the quadrant configuration of the Toothbrush Cluster using the two-temperature \texttt{crossarf$\times$tbabs(bapec+bapec)} model in the 3--8\,keV band. The relative model normalisations of the two components are fixed based on the $\beta$-model analysis as reported in Table\,\ref{double beta contributions}.}
\label{fixed norm from beta model spectral fit}
\centering
\renewcommand{\arraystretch}{1.4}
\begin{tabular}{lcc}
\hline\hline
Parameter (Unit)
& \shortstack{Southern\\component}
& \shortstack{Northern\\component} \\
\hline

$kT$ (keV)
& $8.80^{+0.33}_{-0.32}$
& $6.72^{+0.27}_{-0.28}$ \\

Abundance ($Z/Z_\odot$)
& $0.30^{+0.02}_{-0.02}$
& $0.30$ (coupled) \\

$\sigma_v$ ($\mathrm{km\,s}^{-1}$)
& $170^{+55}_{-40}$
& $255^{+80}_{-80}$ \\
\hline

\multicolumn{3}{l}{\textbf{SE Region}} \\


Redshift
& $0.2224^{+0.0002}_{-0.0006}$
& $0.2222^{+0.0047}_{-0.0097}$ \\

$v_{\rm bulk}$ ($\mathrm{km\,s}^{-1}$)
& $-905^{+50}_{-145}$
& $-955^{+1150}_{-2370}$ \\

Norm ($10^{-3}$)
& $2.17^{+0.07}_{-0.07}$
& $0.26$ (coupled) \\
\hline

\multicolumn{3}{l}{\textbf{NE Region}} \\


Redshift
& $0.2230^{+0.0005}_{-0.0011}$
& $0.2176^{+0.0005}_{-0.0009}$ \\

$v_{\rm bulk}$ ($\mathrm{km\,s}^{-1}$)
& $-760^{+120}_{-270}$
& $-2080^{+120}_{-220}$ \\

Norm ($10^{-3}$)
& $0.93^{+0.04}_{-0.04}$
& $1.02$ (coupled) \\
\hline

\multicolumn{3}{l}{\textbf{NW Region}} \\


Redshift
& $0.2249^{+0.0005}_{-0.0006}$
& $0.2297^{+0.0004}_{-0.0004}$ \\

$v_{\rm bulk}$ ($\mathrm{km\,s}^{-1}$)
& $-295^{+120}_{-145}$
& $880^{+100}_{-100}$ \\

Norm ($10^{-3}$)
& $0.55^{+0.02}_{-0.02}$
& $2.44$ (coupled) \\
\hline

\multicolumn{3}{l}{\textbf{SW Region}} \\


Redshift
& $0.2228^{+0.0003}_{-0.0005}$
& $0.2160^{+0.0019}_{-0.0011}$ \\

$v_{\rm bulk}$ ($\mathrm{km\,s}^{-1}$)
& $-810^{+60}_{-130}$
& $-2470^{+455}_{-270}$ \\

Norm ($10^{-3}$)
& $1.61^{+0.05}_{-0.05}$
& $0.71$ (coupled) \\
\hline

C-stat (Expected C-stat)
& \multicolumn{2}{c}{$4731$ ($4762 \pm 91$)} \\

DOF
& \multicolumn{2}{c}{4173} \\
\hline
\end{tabular}
\end{table}


\end{appendix}


\end{document}